\documentclass[%
reprint,
superscriptaddress,
showpacs,
nofootinbib,
amsmath,amssymb,
prc,
floatfix ]%
{revtex4-1}

\usepackage{color}
\usepackage{CJK}

\usepackage{graphicx}
\usepackage{dcolumn}
\usepackage{bm}
\usepackage{hyperref}

\allowdisplaybreaks

\begin{document}

\begin{CJK*}{UTF8}{}
\title{Microscopic model for the collective enhancement of nuclear level densities}
\CJKfamily{gbsn}
\author{Jie Zhao (赵杰)}%
\affiliation{Center for Quantum Computing, Peng Cheng Laboratory, Shenzhen 518055, China}
\author{Tamara Nik\v{s}i\'c}%
\affiliation{Physics Department, Faculty of Science, University of Zagreb, Bijeni\v{c}ka Cesta 32,
        	      Zagreb 10000, Croatia}             
\author{Dario Vretenar}%
\affiliation{Physics Department, Faculty of Science, University of Zagreb, Bijeni\v{c}ka Cesta 32,
              Zagreb 10000, Croatia}
 \affiliation{ State Key Laboratory of Nuclear Physics and Technology, School of Physics, Peking University, Beijing 100871, China}            

\date{\today}

\begin{abstract}
A microscopic method for calculating nuclear level density (NLD) is developed, based on the framework of energy density functionals. Intrinsic level densities are computed from single-quasiparticle spectra obtained in a finite-temperature self-consistent mean-field (SCMF) calculation that takes into account nuclear deformation, and is specified by the choice of the energy density functional (EDF) and pairing interaction. The total level density is calculated by convoluting the intrinsic density with the corresponding collective level density, determined by the eigenstates of a five-dimensional quadrupole or quadrupole plus octupole collective Hamiltonian. The parameters of the Hamiltonian (inertia parameters, collective potential) are consistently determined by deformation-constrained SCMF calculations using the same EDF and pairing interaction. The model is applied in the calculation of NLD of $^{94,96,98}$Mo, $^{106,108}$Pd, $^{106,112}$Cd, $^{160,162,164}$Dy, $^{166}$Er, and $^{170,172}$Yb, in comparison with available data. It is shown that the collective enhancement of the intrinsic level density, consistently computed from the eigenstates of the corresponding collective Hamiltonian, leads to total NLDs that are in very good agreement with data over the entire energy range of measured values. 
\end{abstract}

\maketitle

\end{CJK*}

\bigskip

\section{Introduction~\label{sec:Introduction}}

Level density is a basic nuclear property that also plays a crucial role in many applications, from calculation of reaction cross sections relevant for nucleosynthesis to energy production. An accurate 
computation of nuclear level density (NLD) is a challenging theoretical task because of the 
complexity of the inter-nucleon interaction and the fact that the number of levels increases exponentially with excitation energy. The most widely used methods for calculating NLDs have been based on the Bethe formula, formulated with
the partition function in the zeroth-order approximation of the Fermi-gas model~\cite{Bethe1937_RMP9-69}. 
In realistic applications to finite nuclei various phenomenological modifications to the original analytical 
formula have been suggested~\cite{Koning2008_NPA810-13,Goriely1996_NPA605-28}.
The extensions of the Bethe formula and its limitations are discussed in Ref.~\cite{Gross1993_PLB318-405}.

A number of microscopic approaches to modelling NLD have been reported, 
such as the Shell-Model Monte Carlo method~\cite{Alhassid1999_PRL83-4265,Alhassid2007_PRL99-162504,Alhassid2015_PRC92-024307}, 
the moments method derived from random matrix theory 
and statistic spectroscopy~\cite{Zelevinsky2018_PLB783-428,Senkov2010_PRC82-024304}, 
the stochastic estimation method~\cite{Shimizu2016_PLB753-13}, 
the Lanczos method using realistic nuclear Hamiltonians~\cite{Ormand2020_PRC102-014315}, 
the self-consistent mean-field approach based on the extended Thomas-Fermi approximation 
with Skyrme forces~\cite{Kolomietz2018_PRC97-064302},
and the exact pairing plus independent particle model at finite
temperature~\cite{Hung2017_PRL118-022502,Dang2017_PRC96-054321,Dey2017_PRC96-054326,Dey2019_PLB789-634}.
Microscopic methods based on the self-consistent Hartree-Fock (HF) plus BCS 
model~\cite{Choudhury1977_PRC16-757,Minato2011_JNST48-984,Demetriou2001_NPA695-95}
and Hartree-Fock-Bogoliubov (HFB) 
model~\cite{Hilaire2006_NPA779-63,Hilaire2001_TEPJA12-169,Goriely2008_PRC78-064307} 
have also been developed to describe NLD. 
In this framework the partition function is determined using the same two-body interaction 
as in the HF plus BCS or HFB mean-field models~\cite{Minato2011_JNST48-984} and, therefore, 
shell, pairing, and deformation effects are included self-consistently.
The intrinsic level density is obtained by an inverse Laplace transform of the partition function with the saddle-point approximation \cite{Bohr1969_Nucl_Structure}.  A collective enhancement of the NLD can be taken into account by a phenomenological or semi-empirical multiplicative factor for rotational and vibrational degrees of freedom \cite{Junghans1998_NPA629-635,Demetriou2001_NPA695-95,Rahmatinejad2020_PRC101-054315,Kargar2007_PRC75-064319,Grimes2019_PRC99-064331}, 
or more microscopically by a combinatorial method using single-particle level schemes obtained in HF plus BCS or HFB calculations~\cite{Hilaire2001_TEPJA12-169,Goriely2008_PRC78-064307}. 
The success of the microscopic self-consistent HFB-based approach to NLDs has also motivated calculations of fission cross sections~\cite{Goriely2009_PRC79-024612,Goriely2011_PRC83-034601},
and studies of nuclear shape evolution in the fission process~\cite{Ward2017_PRC95-024618}. 

In a recent calculation of level densities in Dy and Mo isotopes \cite{Rahmatinejad2020_PRC101-054315}, with single-particle spectra obtained from a Woods-Saxon potential, the collective enhancement of the level densities has been determined using available experimental levels at low excitation energies, and also compared with a phenomenological macroscopic model. However, in many cases, and especially in nuclei far from stability, the phenomenological and semi-empirical approaches cannot be applied on a quantitative level. In this work we develop a microscopic method for calculating NLDs, in which the single-quasiparticle spectrum is obtained using a finite-temperature self-consistent mean-field (SCMF) method, while the collective enhancement is determined from the eigenstates of a corresponding collective Hamiltonian that takes into account quadrupole and octupole degrees of freedom. Both the intrinsic level density and the collective enhancement are determined by the same global energy density functional and pairing interaction. 

For the finite-temperature (FT) and deformation-constrained SCMF calculations we employ the relativistic Hartree-Bogoliubov (RHB) model \cite{Vretenar2005_PR409-101,Meng2006_PPNP57-470,Zhou2016_PS91-063008,Meng2016_WorldSci}. This model has been applied to structure studies over the whole mass table, and 
its beyond-mean-field extension, especially the collective Hamiltonian approach  \cite{Li2016_JPG43-024005}, used in a 
number of calculations of low-energy excitation spectra. Nuclear thermodynamics \cite{Niu2013_PRC88-034308,Zhang2018_PRC97-054302,Zhang2017_PRC96-054308} 
and induced fission dynamics~\cite{Zhao2019_PRC99-054613,Zhao2019_PRC99-014618} 
have also been explored with the FT-RHB model. 

The theoretical framework and methods are introduced in Sec.~\ref{sec:model}.
The details of the calculation and the results for 
$^{94,96,98}$Mo, $^{106,108}$Pd, $^{106,112}$Cd,
$^{160,162,164}$Dy, $^{166}$Er, and $^{170,172}$Yb
are discussed in Sec.~\ref{sec:results}.
Sec.~\ref{sec:summary} contains a short summary of the principal results.

\section{\label{sec:model}Theoretical framework}
Assuming that a nucleus is in a state of thermal equilibrium at temperature $T$, 
it can be described by the finite temperature (FT) Hartree-Fock-Bogoliubov (HFB) theory~\cite{Goodman1981_NPA352-30,Egido1986_NPA451-77}. 
In the grand-canonical ensemble, the expectation value of any operator $\hat{O}$ is given by the ensemble average 
\begin{equation}
\langle \hat{O} \rangle = \textrm{Tr} ~[ \hat{D}\hat{O} ],
\end{equation}
where $\hat{D}$ is the density operator:
\begin{equation}
\hat{D} = {1 \over Z} ~ e^{ -\beta \left( \hat{H}-\lambda \hat{N} \right) }\; .
\end{equation}
$Z$ is the partition function, the inverse temperature 
$\beta=1/k_{B}T$ with the Boltzmann constant $k_{B}$, 
$\hat{H}$ is the Hamiltonian of the system, 
$\lambda$ denotes the chemical potential, and $\hat{N}$ is the particle number operator. 
The entropy of the compound nuclear system is $S=-k_{B} \langle \hat{D} \ln \hat{D} \rangle$.
In this work we employ the multidimensionally-constrained (MDC) RHB
model~\cite{Lu2012_PRC85-011301R,Lu2014_PRC89-014323,Zhou2016_PS91-063008,Zhao2016_PRC93-044315} 
at finite temperature to calculate the single-nucleon quasiparticle states. 
The minimization of the grand-canonical potential 
$\Omega = \langle \hat{H} \rangle + TS - \mu \langle \hat{N} \rangle$, where $\mu=\beta\lambda$,
yields the FT-RHB equation 
\begin{equation}
\label{eq:rhb}
 \int d^{3}\bm{r}^{\prime}
  \left(\begin{array}{cc} h-\lambda & \Delta \\ -\Delta^{*} & -h+\lambda \end{array}\right)
  \left(\begin{array}{c} U_{k} \\ V_{k} \end{array}\right)
 = E_{k}\left(\begin{array}{c} U_{k} \\ V_{k} \end{array}\right)\;.
\end{equation}
$E_{k}$ is the quasiparticle energy and $\hat{h}$ denotes the single-particle Hamiltonian
\begin{equation}
\label{eq:hamiltonian}
 \hat{h} = \bm{\alpha} \cdot \bm{p} + \beta[M+S(\bm{r})] + V_{0}(\bm{r}) + \Sigma_R(\bm{r}),
\end{equation}
where, for the relativistic energy-density functional DD-PC1~\cite{Niksic2008_PRC78-034318}, 
the scalar potential, vector potential, and rearrangement terms read
\begin{eqnarray}
       S & = & \alpha_{S}(\rho) \rho_{S} + \delta_{S}\triangle\rho_{S}, \nonumber \\
 V_{0} & = & \alpha_{V}(\rho) \rho_{V} + \alpha_{TV}(\rho) \vec{\rho}_{TV} \cdot \vec{\tau} 
             + e{1-\tau_3 \over 2} A_{0}, \nonumber \\
\Sigma_R & = & {1\over2} {\partial \alpha_{S} \over \partial \rho} \rho_{S}^2
	     + {1\over2} {\partial \alpha_{V} \over \partial \rho} \rho_{V}^2
	     + {1\over2} {\partial \alpha_{TV} \over \partial \rho} \rho_{TV}^2\; ,
\end{eqnarray}
respectively. $M$ is the nucleon mass, $\alpha_{S}({\rho})$, $\alpha_{V}({\rho})$,  and $\alpha_{TV}({\rho})$ are
density-dependent couplings for different space-isospace channels, 
$\delta_{S}$ is the coupling constant of the derivative term,
and $e$ is the electric charge. 
The single-nucleon densities 
$\rho_{S}$ (scalar-isoscalar density), $\rho_{V}$ (time-like component of the 
isoscalar current), and $\rho_{TV}$  (time-like component of the isovector current), are defined by the following relations:
\begin{eqnarray}
\label{eq:densities}
 \rho_{S}(\bm{r})  & = & \sum_{k>0} V_{k}^{\dagger}(\bm{r}) \gamma_{0} (1-f_{k}) V_{k}(\bm{r}) 
 	+ U_{k}^{T}(\bm{r}) \gamma_{0} f_{k} U_{k}^{*}(\bm{r}), \nonumber \\
 \rho_{TV}(\bm{r}) & = & \sum_{k>0} V_{k}^{\dagger}(\bm{r}) \tau_{3} (1-f_{k}) V_{k}(\bm{r})
 	+ U_{k}^{T}(\bm{r}) \tau_{3} f_{k} U_{k}^{*}(\bm{r}), \nonumber \\
 \rho_{V}(\bm{r})  & = & \sum_{k>0} V_{k}^{\dagger}(\bm{r}) (1-f_{k}) V_{k}(\bm{r})
 	+ U_{k}^{T}(\bm{r}) f_{k} U_{k}^{*}(\bm{r}),
\end{eqnarray}
where $f_k$ is the thermal occupation probability of the quasiparticle state $k$
\begin{equation}
f_{k} = {1 \over 1+e^{\beta E_k}},
\end{equation}
The pairing potential reads
\begin{equation}
\begin{aligned}
 \Delta (\bm{r}_{1}\sigma_{1},\bm{r}_{2}\sigma_{2})
 = & \int d^{3}\bm{r}_{1}^{\prime} d^{3}\bm{r}_{2}^{\prime} 
      \sum_{\sigma_{1}^{\prime}\sigma_{2}^{\prime}} \\
   & V^{{\rm pp}} 
      (\bm{r}_{1}\sigma_{1}, \bm{r}_{2}\sigma_{2},
       \bm{r}_{1}^{\prime}\sigma_{1}^{\prime}, \bm{r}_{2}^{\prime}\sigma_{2}^{\prime}) \\
  & \times \kappa
      (\bm{r}_{1}^{\prime}\sigma_{1}^{\prime}, 
       \bm{r}_{2}^{\prime}\sigma_{2}^{\prime}),
\end{aligned}
\end{equation}
where $V^{{\rm pp}}$ is the effective pairing interaction
and $\kappa$ is the pairing tensor,
\begin{equation}
 \kappa
 = \sum_{k>0} V^{*}_{k} (1-f_{k}) U_{k}  + U_{k} f_{k} V_{k}^{\dagger}.
\end{equation}
In the particle-particle channel we use a separable pairing force of finite range \cite{Tian2009_PLB676-44}:
\begin{equation}
V(\mathbf{r}_1,\mathbf{r}_2,\mathbf{r}_1^\prime,\mathbf{r}_2^\prime) = G_0 ~\delta(\mathbf{R}-
\mathbf{R}^\prime) P (\mathbf{r}) P(\mathbf{r}^\prime) \frac{1}{2} \left(1-P^\sigma\right),
\label{pairing}
\end{equation}
where $\mathbf{R} = (\mathbf{r}_1+\mathbf{r}_2)/2$ and $\mathbf{r}=\mathbf{r}_1- \mathbf{r}_2$
denote the center-of-mass and the relative coordinates, respectively. $P(\mathbf{r})$ reads 
\begin{equation}
P(\mathbf{r})=\frac{1}{\left(4\pi a^2\right)^{3/2}} e^{-\mathbf{r}^2/4a^2}.
\end{equation}
The two parameters of the interaction were originally 
adjusted to reproduce the density dependence of the pairing gap in nuclear matter at the
Fermi surface calculated with the D1S parameterization of the Gogny force~\cite{Berger1991_CPC63-365}.

The entropy is computed using the relation:
\begin{equation}
S = -k_{B} \sum_{k} \left[ f_{k} \ln f_{k} + (1 - f_{k}) \ln (1 - f_{k}) \right].
\end{equation}
Employing the saddle point approximation \cite{Bohr1969_Nucl_Structure}, one obtains the following expression for the intrinsic level density $\rho_{i}$ 
\begin{equation}
\label{eq:ld}
\rho_{i} = \frac{e^{S}}{(2\pi)^{3/2}D^{1/2}},
\end{equation}
where $D$ is the determinant of a $3 \times 3$ matrix that contains the second derivatives of the entropy with respect to 
$\beta$ and $\mu_\tau = \beta \lambda_\tau$ ($\tau \equiv p,n$) at the saddle point.
The intrinsic excitation energy is calculated as $U_{i}(T) = E(T) - E(0)$, with 
$E(T)$ the binding energy of the nucleus at temperature $T$.

With the assumption of a decoupling between intrinsic and collective degrees of freedom, 
the excitation energy of a nucleus can be written as $U = U_{i} + U_{c}$, where $U_{c}$ is the collective 
excitation energy~\cite{Rahmatinejad2020_PRC101-054315}.
The total level density is obtained as 
\begin{equation}
\label{eq:LDtot}
\rho_{\rm{tot}}(U) = \int \rho_{i}(U_{i}) \rho_{c}(U-U_{i}) dU_{i},
\end{equation}
with the collective level density 
\begin{equation}
\label{eq:LDcoll}
\rho_{c}(U) = \sum_{c} \delta(U-U_{c}) \tau_{c}(U_{c}).
\end{equation}
For a collective state with the angular momentum $I_{c}$, the degeneracy 
is $\tau_{c}(U_{c}) = 2 I_{c} + 1$.

In the microscopic model used in the present work the collective levels are eigenstates either of 
the five-dimensional quadrupole~\cite{Niksic2009_PRC79-034303}
or the axial quadrupole-octupole Hamiltonians~\cite{Li2013_PLB726-866}.  
In the former case in which one considers only quadrupole degrees of freedom,
the collective Hamiltonian reads
\begin{equation}
\begin{aligned}
\label{eq:CHtri}
\hat{H}_{\rm{coll}} (\beta,\gamma) =& -\frac{\hbar^{2}}{2\sqrt{wr}} \left\{  \frac{1}{\beta^{4}} 
	\left[ \frac{\partial}{\partial \beta} \sqrt{\frac{r}{w}} \beta^{4} 
	B_{\gamma\gamma} \frac{\partial}{\partial \beta} \right.\right. \\
	& \left. - \frac{\partial}{\partial \beta} \sqrt{\frac{r}{w}} \beta^{3} 
	B_{\beta\gamma} \frac{\partial}{\partial \gamma}  \right]  \\
	&+ \frac{1}{\beta\sin 3\gamma} \left[ -\frac{\partial}{\partial \gamma} 
	\sqrt{\frac{r}{w}} \sin 3\gamma B_{\beta\gamma} \frac{\partial}{\partial \beta} \right. \\
	& \left.\left. + \frac{1}{\beta} \frac{\partial}{\partial \gamma} \sqrt{\frac{r}{w}} 
	\sin 3\gamma B_{\beta\beta} \frac{\partial}{\partial \gamma} \right] \right\} \\
	&+ \frac{1}{2} \sum_{k=1}^{3} \frac{\hat{J}_{k}^2}{\mathcal{I}_{k}}
	+ V(\beta,\gamma),
\end{aligned}
\end{equation}
where $B_{\beta\beta}$, $B_{\beta\gamma}$, $B_{\gamma\gamma}$ are the mass parameters, 
$\mathcal{I}_{k}$ is the moment of inertia, $V(\beta,\gamma)$ denotes the collective potential,
$w$ and $r$ are functions of the mass parameters and moments of inertia.

In the case that includes octupole correlations, the current implementation of the collective Hamiltonian is restricted to axial symmetry, that is, only the axial quadrupole and octupole deformations are considered as collective coordinates. This approximation is justified in heavy, axially deformed nuclei that will be examined in this work. The axial quadrupole-octupole collective Hamiltonian takes the form
\begin{equation}
\begin{aligned}
\label{eq:CHoct}
\hat{H}_{\rm{coll}} (\beta_{2},\beta_{3}) =& -\frac{\hbar^{2}}{2\sqrt{w\mathcal{I}}}  
	\left[ \frac{\partial}{\partial \beta_{2}} \sqrt{\frac{\mathcal{I}}{w}}
	B_{33} \frac{\partial}{\partial \beta_{2}} \right. \\
	& - \frac{\partial}{\partial \beta_{2}} \sqrt{\frac{\mathcal{I}}{w}} 
	B_{23} \frac{\partial}{\partial \beta_{3}}  
	- \frac{\partial}{\partial \beta_{3}} \sqrt{\frac{\mathcal{I}}{w}} B_{23} \frac{\partial}{\partial \beta_{2}} \\
	& \left. + \frac{\partial}{\partial \beta_{3}} \sqrt{\frac{\mathcal{I}}{w}} 
	B_{22} \frac{\partial}{\partial \beta_{3}} \right]
	+ \frac{\hat{J}^2}{2\mathcal{I}}
	+ V(\beta_{2},\beta_{3}).
\end{aligned}
\end{equation}
The mass parameters $B_{22}$, $B_{23}$, $B_{33}$, and the moment of inertia $\mathcal{I}$ are functions 
of the quadrupole $\beta_{2}$ and octupole $\beta_{3}$ deformations. $w=B_{22}B_{33} - B_{23}^{2}$.

The mass parameters, moments of inertia, and collective potentials as functions of the collective coordinates ${\bf q} \equiv (\beta, \gamma)$ or $(\beta_2, \beta_3)$, are completely determined by the defomation-constrained self-consistent RHB calculations at zero temperature for 
a specific choice of the nuclear energy density functional and pairing interaction. 
In the persent version of the model, the mass parameters defined as the inverse of the mass tensor 
$B_{ij} ({\bf q}) = {\cal M} ^{-1}_{ij} ({\bf q})$,  
are calculated in the perturbative cranking approximation~\cite{Zhao2015_PRC92-064315}
 \begin{equation}
\label{eq:pmass}
{\cal M}^{C_p} = \hbar^2 {\it M}_{(1)}^{-1} {\it M}_{(3)} {\it M}_{(1)}^{-1}, 
\end{equation}
 where 
\begin{equation}
\label{eq:mmatrix}
\left[ {\it M}_{(k)} \right]_{ij} = \sum_{\mu\nu} 
    {\left\langle 0 \left| \hat{Q}_i \right| \mu\nu \right\rangle
     \left\langle \mu\nu \left| \hat{Q}_j \right| 0 \right\rangle
     \over (E_\mu + E_\nu)^k}\; . 
\end{equation}
$|\mu\nu\rangle$ are two-quasiparticle 
wave functions, and $E_\mu$ and $E_\nu$ the corresponding quasiparticle energies. $\hat{Q}_i$
denotes the multipole operators that correspond to the collective degrees of freedom.
The collective potential $V$ is obtained by subtracting the 
vibrational zero-point energy (ZPE) from the total RHB 
deformation energy. Following the prescription of 
Refs.~\cite{Sadhukhan2013_PRC88-064314,Sadhukhan2014_PRC90-061304,Baran2007_IJMPE16-443,Staszczak2013_PRC87-024320}, 
the ZPE is computed using the Gaussian overlap approximation, 
\begin{equation}
\label{eq:zpe}
E_{\rm ZPE} = {1\over4} {\rm Tr} \left[ {\it M}_{(2)}^{-1} {\it M}_{(1)} \right].
\end{equation}
The microscopic self-consistent solutions of the constrained RHB equations, that is, the 
single-quasiparticle energies and wave functions on the entire energy surface as functions of the 
deformations, provide the microscopic input for the calculation of both the 
collective inertia and zero-point energy. 
The Inglis-Belyaev formula is used for the rotational moment of inertia.
From the diagonalization of the collective Hamiltonian one obtains the collective energy spectrum.  

The deformation-dependent energy landscape is mapped in a self-consistent RHB 
calculation with constraints on the mass multipole moments $Q_{\lambda\mu} = r^\lambda Y_{\lambda \mu}$.
The nuclear shape is parameterized by the deformation parameters
\begin{equation}
 \beta_{\lambda\mu} = {4\pi \over 3AR^\lambda} \langle Q_{\lambda\mu} \rangle.
\end{equation}
The self-consistent RHB equations are solved by expanding the single-nucleon spinors in a
harmonic oscillator (HO) basis. The present calculations have been performed 
in a HO basis truncated to $N_f = 20$ oscillator shells
for the axially symmetric case (heavy Dy, Er, and Yb nuclei), while $N_{f} = 16$ has been used for the triaxial case (medium-heavy Mo, Pd, and Cd isotopes).
For details of the MDC-RHB model we refer the reader to 
Refs.~\cite{Lu2014_PRC89-014323,Zhao2017_PRC95-014320}.

\section{\label{sec:results}Illustrative calculations in the mass A=100 and A=160 regions}
The microscopic approach and the particular model developed in this work will be illustrated 
with calculations of the total level densities for  
$^{94,96,98}$Mo, $^{106,108}$Pd, $^{106,112}$Cd, and
$^{160,162,164}$Dy, $^{166}$Er, $^{170,172}$Yb. 
The relativistic energy density functional DD-PC1~\cite{Niksic2008_PRC78-034318} is used in the particle-hole channel, while 
particle-particle correlations are described by the separable pairing force (\ref{pairing}) in the Bogoliubov approximation. 

\begin{figure}[!]
\flushleft
 \includegraphics[width=0.22\textwidth]{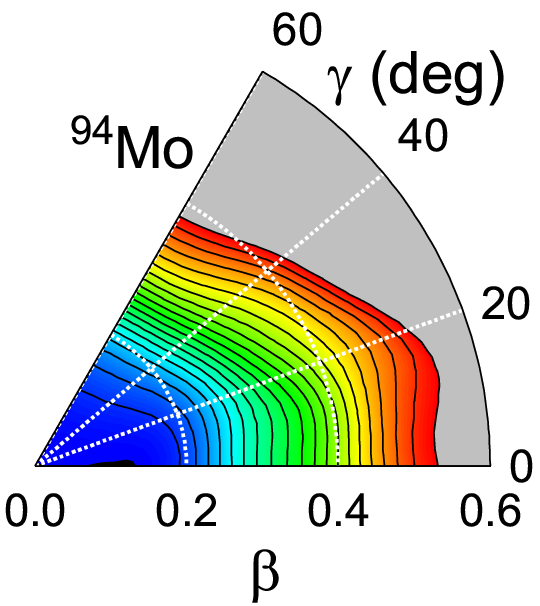}~~
 \includegraphics[width=0.22\textwidth]{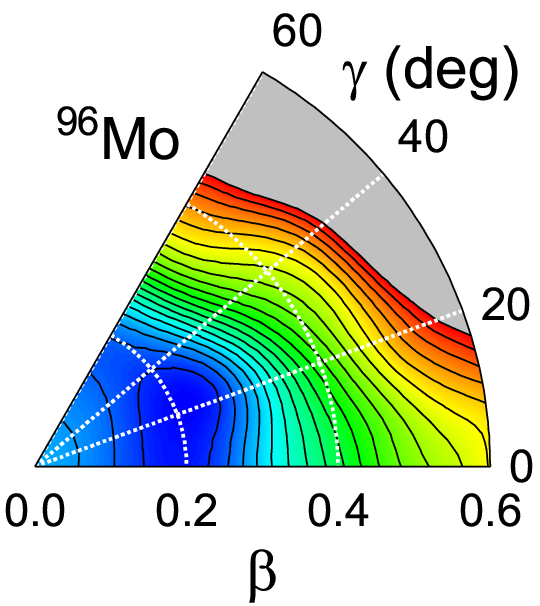} \\
 \includegraphics[width=0.22\textwidth]{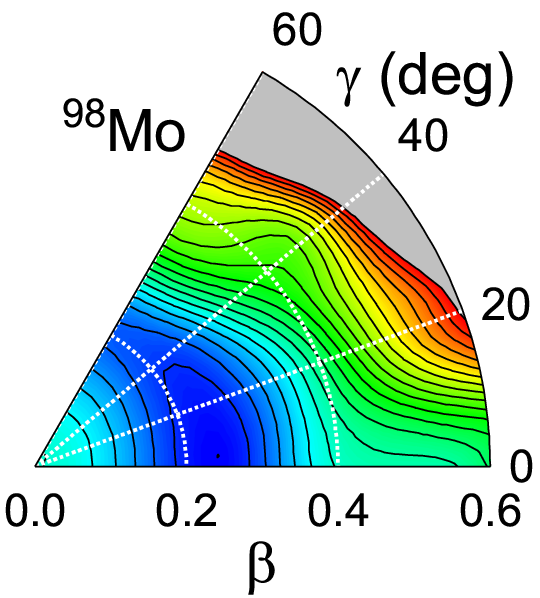}~~
 \includegraphics[width=0.22\textwidth]{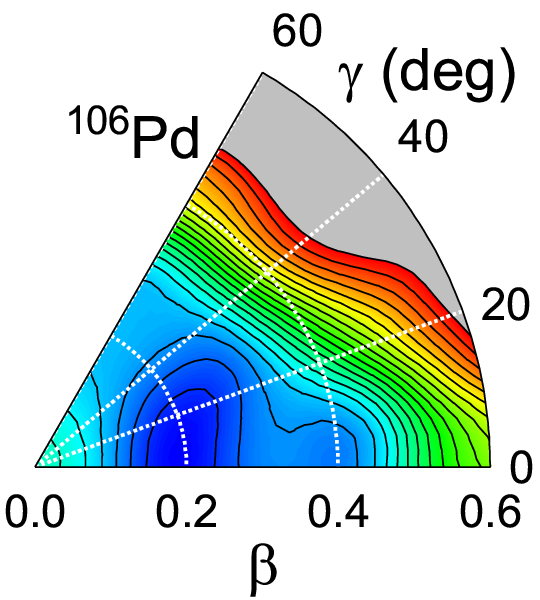} \\
 \includegraphics[width=0.22\textwidth]{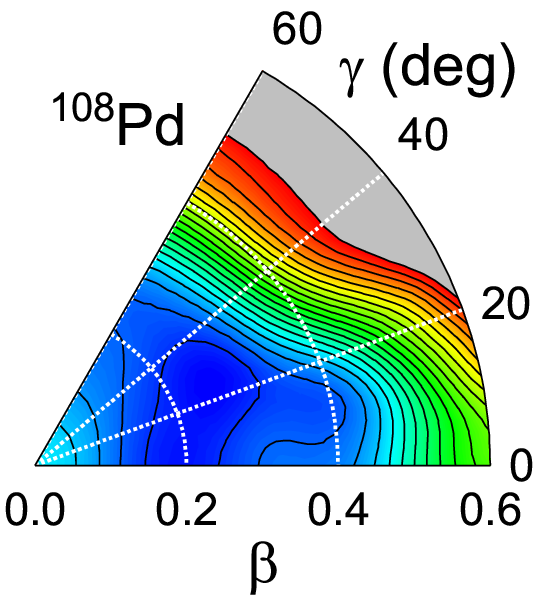}~~
 \includegraphics[width=0.22\textwidth]{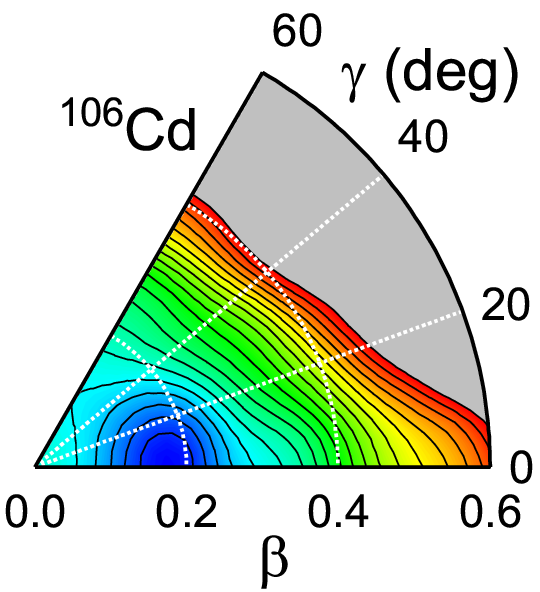} \\
 \includegraphics[width=0.22\textwidth]{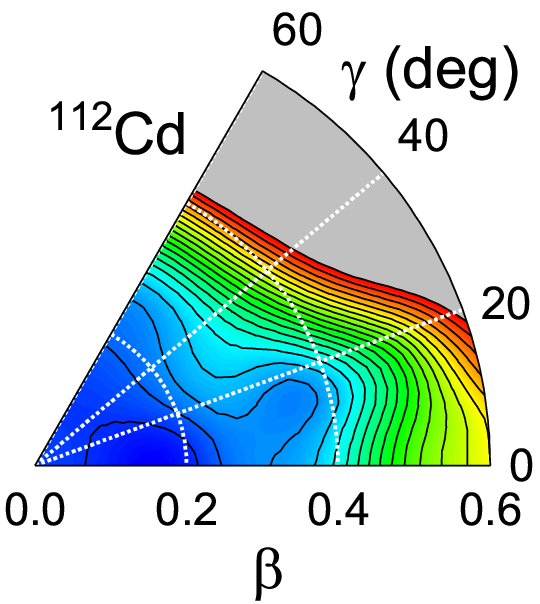}
\caption{(Color online)~\label{fig:PES_tri}%
Self-consistent triaxial quadrupole deformation-constrained energy surfaces of 
$^{94,96,98}$Mo, $^{106,108}$Pd, and $^{106,112}$Cd
in the $\beta$-$\gamma$ plane $(0 \leq \gamma \leq 60^{\circ})$.
For each nucleus the energies are normalized with respect to the binding energy of the global minimum.
The contours join points on the surface with the same energy, 
and the spacing between neighbouring contours is 0.5 MeV.
}
\end{figure}

In the first step, for each nucleus a FT-RHB calculation is performed at 
the equilibrium (global) minimum to determine the intrinsic level density. 
\begin{figure}[!]
\flushleft
 \includegraphics[width=0.48\textwidth]{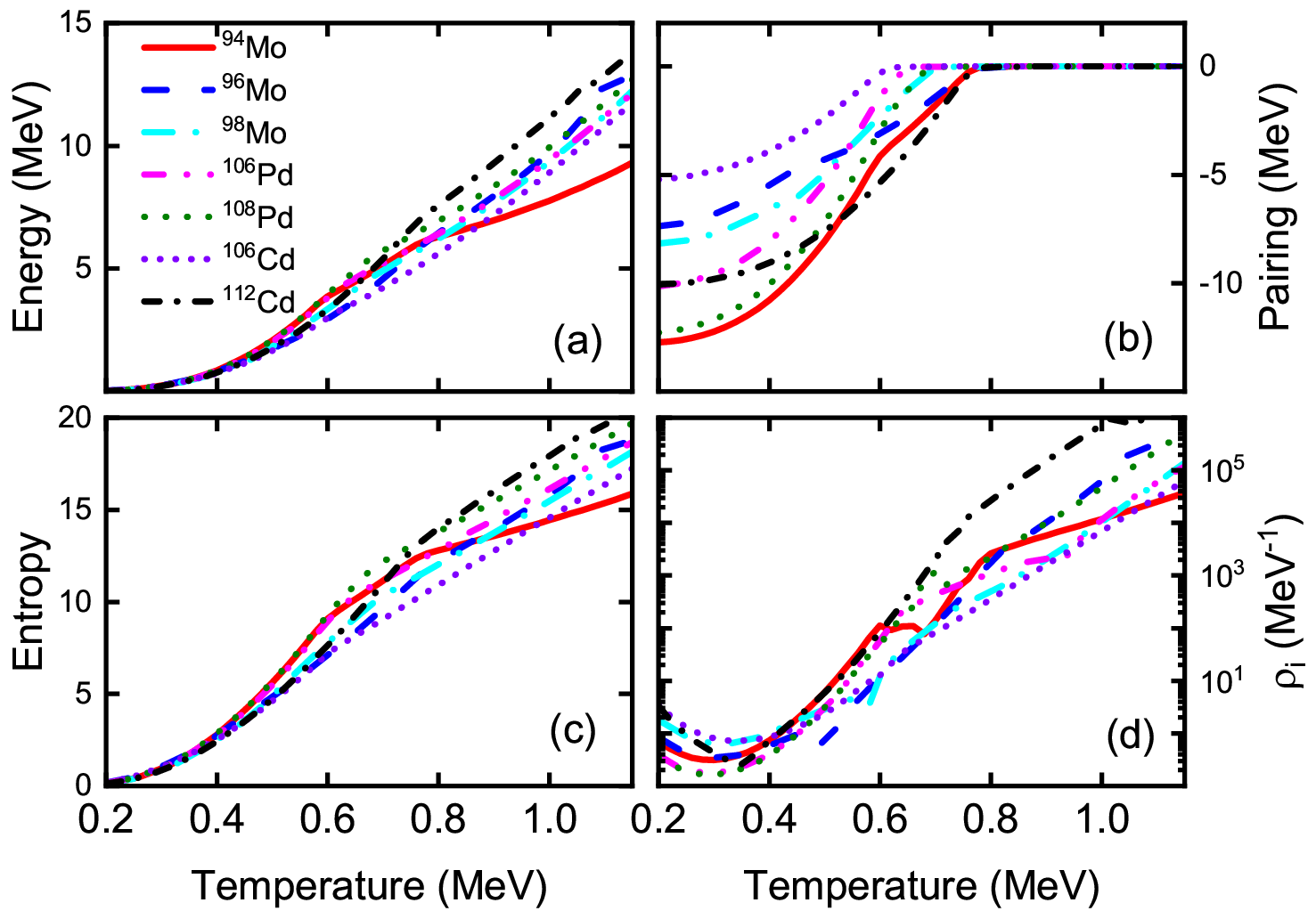}
 \caption{(Color online)~\label{fig:Obs_tri}%
The energy of the equilibrium (global) minimum (a), the pairing energy (b), entropy (c), and intrinsic level density $\rho_{i}$ (d),  as functions of temperature for $^{94,96,98}$Mo, $^{106,108}$Pd, and $^{106,112}$Cd. The results are obtained in finite-temperature triaxial RHB calculations with the DD-PC1 energy density functional and finite-range pairing interaction, as described in the previous section. 
}
\end{figure}
In Fig.~\ref{fig:Obs_tri} we display the calculated energies of the equilibrium minima, the pairing energies, entropies, and intrinsic level densities as functions of temperature for 
$^{94,96,98}$Mo, $^{106,108}$Pd, $^{106,112}$Cd. Most of these nuclei exhibit deformation 
energy surfaces that are soft in $\gamma$ deformation (cf. Fig.~\ref{fig:PES_tri}), while the octupole deformation $\beta_{3}$ does not play a significant role at low energies. The RHB calculation has, therefore, been restricted to triaxial quadrupole deformations. 

As shown in Fig.~\ref{fig:Obs_tri} (b), pairing correlations decrease rapidly as temperature increases and the pairing energy vanishes at the critical temperature $T_{c}=0.6 \sim 0.8$ MeV.
In the Fermi-gas model the intrinsic excitation energy can be approximated by the Bethe 
formula $U_{i} = aT^{2}$, where $a$ is the level density parameter. 
From Fig.~\ref{fig:Obs_tri} (a) one notices that the energy indeed increases quadratically with temperature, and a change of slope can be associated with the pairing phase transition at the critical temperature. Fig.~\ref{fig:Obs_tri} (c) shows that below the critical temperature $T_{c}$ the entropy 
also increases quadratically with temperature.
After the pairing phase transition the entropy increases linearly with $T$, 
in agreement with the Bethe formula $S=2aT$.  
The intrinsic level density increases exponentially with the entropy (cf. Eq.~(\ref{eq:ld})),  
and a change of slope, or even a discontinuity, is found around $T_{c}$, as shown Fig.~\ref{fig:Obs_tri} (d).

In the second step a large scale zero-temperature MDC-RHB calculation is performed to generate the 
collective potential energy surface (PES), single-quasiparticle energies and wave functions in the $(\beta,\gamma)$ plane. 
Fig.~\ref{fig:PES_tri} displays the resulting deformation energy surfaces of 
$^{94,96,98}$Mo, $^{106,108}$Pd, $^{106,112}$Cd.
At zero temperature the ground state shape for $^{94}$Mo is almost spherical
and the PES is soft in both $\beta$ and $\gamma$ directions. 
The equilibrium deformation of $^{96}$Mo is at $(\beta, \gamma) \sim (0.2, 21^{\circ})$, 
and $(\beta, \gamma) \sim (0.2, 13^{\circ})$ for $^{108}$Pd. 
The isotopes $^{98}$Mo, $^{106}$Pd, $^{106,112}$Cd exhibit $\beta$-deformed minima at  
$\beta = 0.15 \sim 0.25$. 
As noted above, the PESs for all these nuclei are rather soft in the $\gamma$ direction. 
With the single-quasiparticle energies and wave functions determined in self-consistent RHB calculations, the corresponding mass parameters, 
moments of inertia, and ZPE over the entire PES can be computed. These quantities specify the 
collective Hamiltonian Eq.~(\ref{eq:CHtri}).

\begin{figure}[!]
\flushleft
 \includegraphics[width=0.48\textwidth]{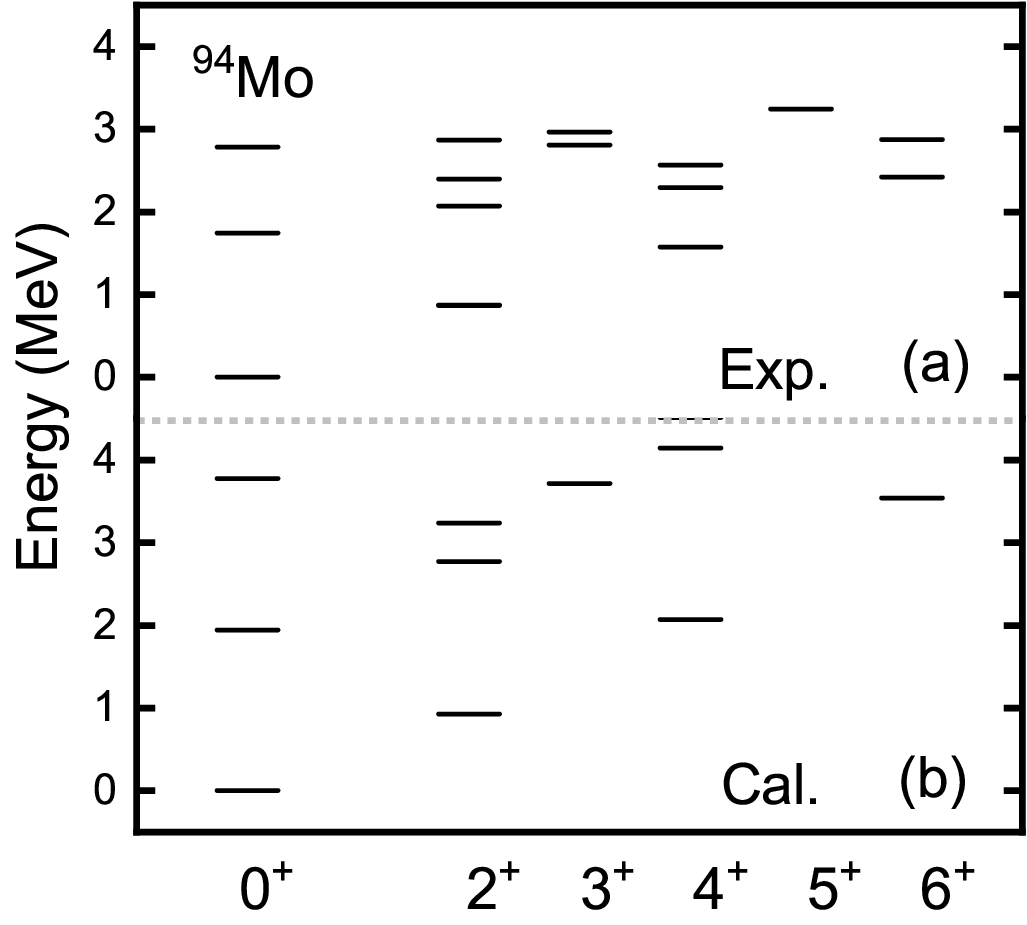}
 \caption{(Color online)~\label{fig:Mo94_cLevl}%
 The calculated positive-parity low-spin states of $^{94}$Mo (b) and their possible experimental counterparts (a). Not all known levels 
 of these spins are shown in panel (a). For a detailed discussion see Ref.~\cite{Fransen2003_PRC67-024307}.
}
\end{figure}
To illustrate the level of agreement with low-energy experimental levels, 
in Fig.~\ref{fig:Mo94_cLevl} we compare the calculated low-spin collective levels of $^{94}$Mo
with the available data from Ref.~\cite{Fransen2003_PRC67-024307}.
The experimental levels are shown in the upper panel, and the eigenstates of the quadrupole 
collective Hamiltonian in the lower panel. The calculated levels are in good qualitative agreement with experiment, except for the fact that the calculated excitation spectrum is somewhat stretched out compared to data. In particular, the moment of inertia of the theoretical yrast band is smaller than the empirical one. This is because the collective inertia is calculated from the Inglis-Belyaev formula which does not include Thouless-Valatin rearrangement contributions and, therefore, predicts effective moments of inertia that are smaller than empirical values.  
The predicted energies of $2_{1}^{+}$ and $0_{2}^{+}$ are 0.92 MeV and 1.94 MeV, respectively,
are compared to the experimental values: 0.87 MeV and 1.74 MeV.
The predicted energy of $4_{1}^{+}$ is 2.07 MeV, which is again considerably above the experimental  value of 1.57 MeV. Here we note that, while the theoretical states are purely collective, there are indications of 
 non-collective components in the $4_{1}^{+}$ state~\cite{Mu2018_SCPMA61-012011}.
For some levels at higher energies, for instance, the experimental values for $0_{3}^{+}$ and $6_{1}^{+}$ are 2.78 MeV and 2.87 MeV, respectively, while the calculation gives the values of 3.78 MeV and 3.54 MeV. In addition to the perturbative cranking approximation used to calculate the mass parameters, we also note that, in particular for the excited $0^{+}$ states, another effect that is not included in the model is the coupling of nuclear shape oscillations with pairing vibrations, that is, vibration of the pairing density. However, the aim of the present study is not a detailed reproduction of the low-energy spectra and, therefore, the qualitative level of agreement between model calculations and experiment, illustrated in Fig.~\ref{fig:Mo94_cLevl}, should be sufficient for an estimate of the collective enhancement of the level density.
\begin{figure}[!]
 \includegraphics[width=0.48\textwidth]{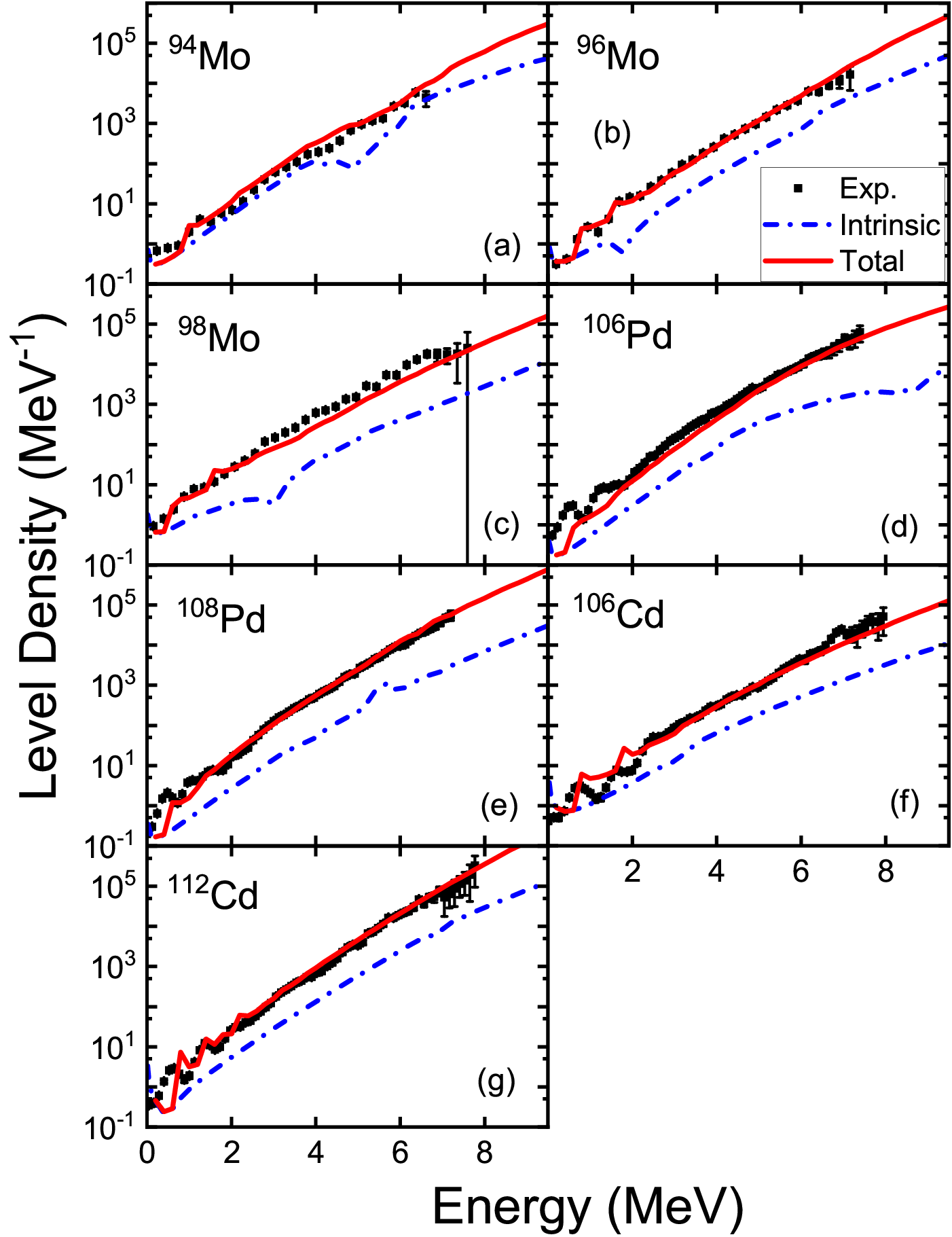}
\caption{(Color online)~\label{fig:LD_tri}%
The calculated intrinsic level densities (dash-dotted blue) and total level densities (solid red),  
as functions of excitation energy for $^{94,96,98}$Mo, $^{106,108}$Pd, and $^{106,112}$Cd.
The data (black squares) are from Refs.~\cite{Utsunomiya2013_PRC88-015805,Eriksen2014_PRC90-044311,Larsen2013_PRC87-014319}.  
}
\end{figure}

\begin{figure}[!]
 \includegraphics[width=0.4\textwidth]{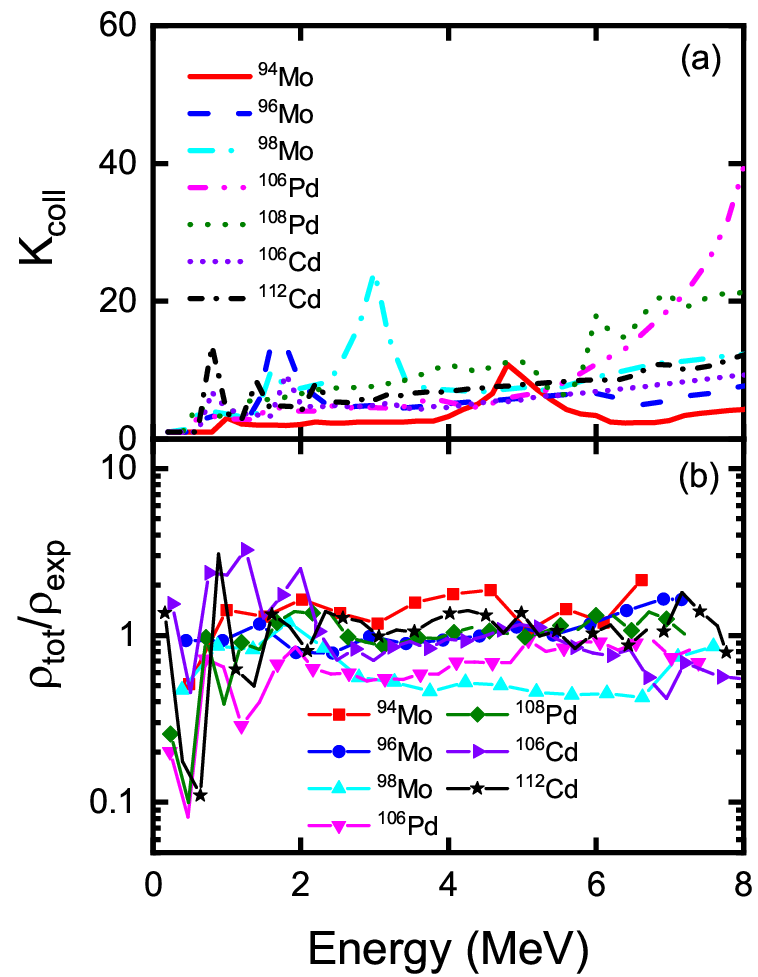}
\caption{(Color online)~\label{fig:Kcoll_tri}%
The collective enhancement factors $K_{\rm{coll}}$ (a), and ratios $\rho_{\rm{tot}} / \rho_{\rm{exp}}$ of calculated and experimental 
level densities (b), as functions of excitation energy for $^{94,96,98}$Mo, $^{106,108}$Pd, and $^{106,112}$Cd. 
}
\end{figure}

Employing the collective levels obtained by diagonalization of the quadrupole Hamiltonian Eq.~(\ref{eq:CHtri}), 
the total level densities can now be computed from Eqs.~(\ref{eq:LDtot}) and (\ref{eq:LDcoll}). In 
Fig.~\ref{fig:LD_tri} we compare, for $^{94,96,98}$Mo, $^{106,108}$Pd, and $^{106,112}$Cd, 
the intrinsic level densities calculated with the FT-RHB model 
Eq.~(\ref{eq:ld}) (dash-dotted blue) and the corresponding total level densities (solid red), with the available data below  $\lesssim 8$ MeV \cite{Utsunomiya2013_PRC88-015805,Eriksen2014_PRC90-044311,Larsen2013_PRC87-014319}.
Obviously the intrinsic level densities cannot reproduce the data in any of these nuclei, and clearly indicate the necessity for including additional degrees of freedom. The consistent inclusion of collective enhancement brings the total theoretical level densities in agreement with data over the whole interval of experimentally determined values.

By comparing the total and intrinsic level densities, in the upper panel of Fig.~\ref{fig:Kcoll_tri} we plot the collective 
enhancement factors 
\begin{equation}
\rho_{\rm{tot}}(U) = K_{\rm{coll}} (U) \rho_{\rm{i}}(U)\; ,
\end{equation}
as functions of excitation energy for $^{94,96,98}$Mo, $^{106,108}$Pd, and $^{106,112}$Cd.  In general, $K_{\rm{coll}} (U)$ exhibits an increase with energy 
in the interval below $\lesssim 8$ MeV. The pronounced peaks at $\approx 5$ MeV for $^{94}$Mo, and at $\approx 3$ Mev for $^{98}$Mo are actually caused by the dips of the intrinsic level densities (cf. Fig.~\ref{fig:LD_tri}), and can be related to a collapse of pairing correlations at these energies. This is an artefact of the SCMF calculation that does include projection on good particle number and is, therefore, unphysical. In the lower panel we plot the 
ratios $\rho_{\rm{tot}} / \rho_{\rm{exp}}$. Except for the oscillations at very low energies below 1 MeV where there are only a few levels,  for most of these nuclei the ratio is actually close to 1 over the entire low-energy interval. 
\begin{figure}
\flushleft
 \includegraphics[width=0.48\textwidth]{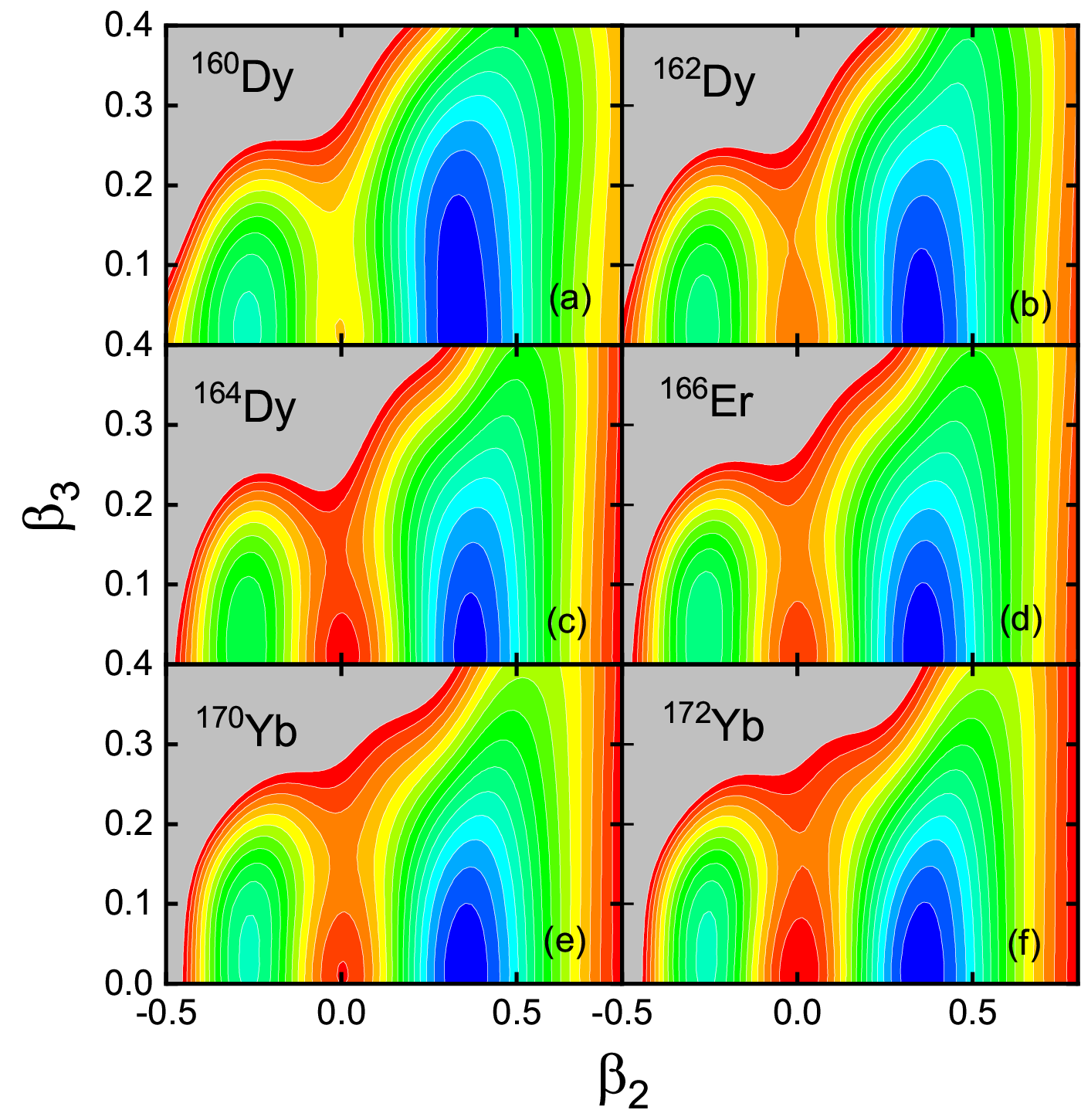}
 \caption{(Color online)~\label{fig:PES_oct}%
Self-consistent RHB axially symmetric deformation energy surfaces of 
$^{160,162,164}$Dy, $^{166}$Er, and $^{170,172}$Yb in the $(\beta_{2}, \beta_{3})$ plane.
For each nucleus the energies are normalized with respect to the binding energy of the global minimum.
The contours join points on the surface with the same energy, 
and the spacing between neighbouring contours is 1.0 MeV.
}
\end{figure}

Several studies based on the non-relativistic Gogny HFB and relativistic RHB models have shown that heavier nuclei in the mass $A \approx 160-170$ region, such as Dy, Er and Yb isotopes, 
exhibit axially symmetric equilibrium shapes, but their potential energy surfaces are rather soft in the octupole $\beta_{3}$ direction.  This is illustrated in Fig.~\ref{fig:PES_oct}, where we display the two-dimensional RHB deformation energy surfaces of $^{160,162,164}$Dy, $^{166}$Er, and $^{170,172}$Yb in the $(\beta_{2}, \beta_{3})$ plane calculated at zero temperature. One notices that, although the global minima are located at $\beta_{2} = 0.3 \sim 0.4$ and $\beta_{3} = 0$, the minima 
are extended in the direction of axial octupole deformation $\beta_{3}$. For this reason we expect a significant contribution of octupole vibrations to the low-energy collective states. As the current implementation of our collective Hamiltonian does not allow the simultaneous breaking of axial and reflection symmetries, in this case we will employ the axially symmetric and reflection asymmetric quadrupole-octupole Hamiltonian of Eq.~(\ref{eq:CHoct}) to calculate the collective enhancement of the RHB intrinsic level densities. 

\begin{figure}[!]
\flushleft
 \includegraphics[width=0.48\textwidth]{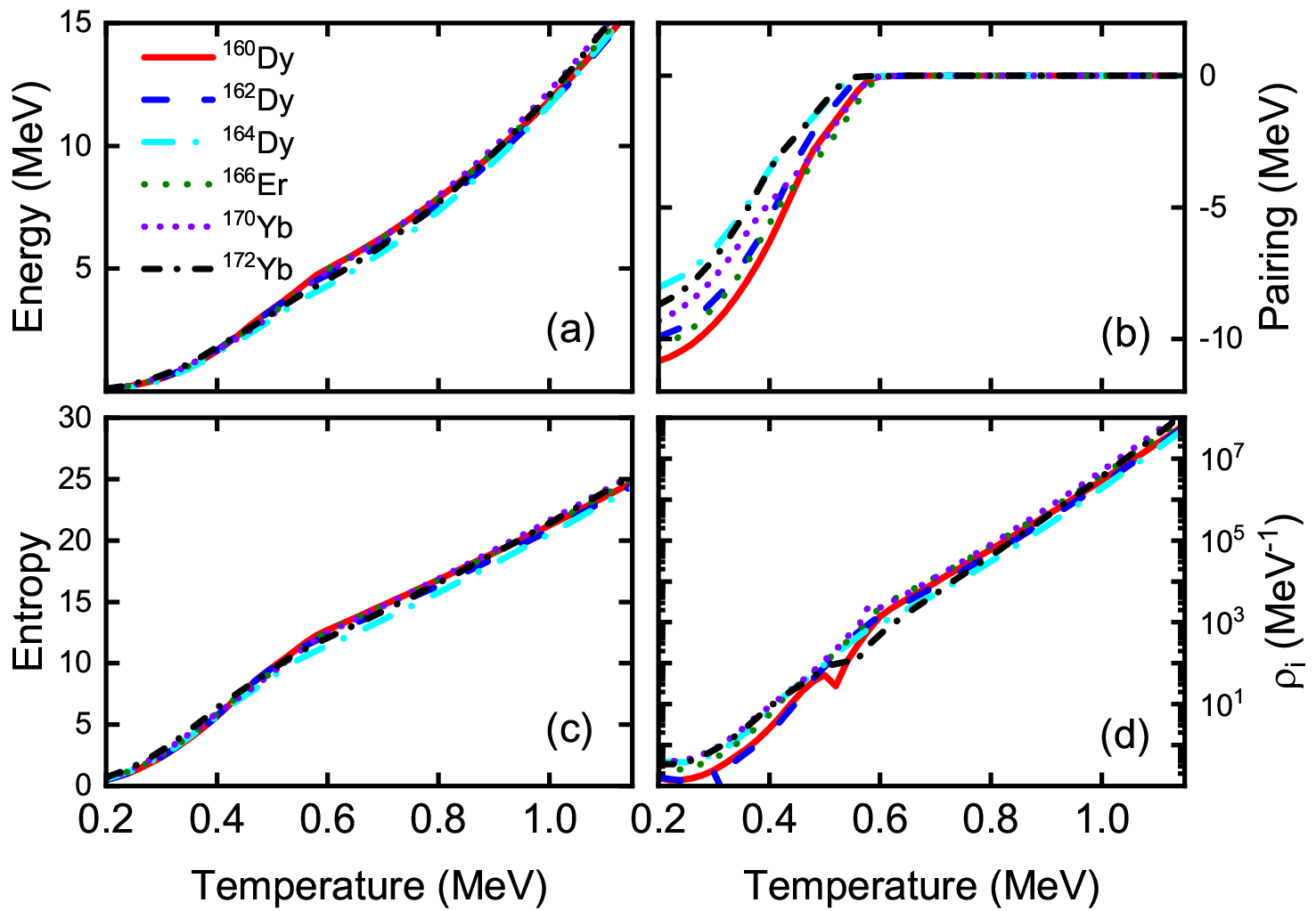}
 \caption{(Color online)~\label{fig:Obs_oct}%
 Same as in the caption to Fig.~\ref{fig:Obs_tri} but for the axially-symmetric and reflection-asymmetric 
 RHB calculations of $^{160,162,164}$Dy, $^{166}$Er, and $^{170,172}$Yb. 
}
\end{figure}

Axially symmetric and reflection asymmetric FT-RHB calculations are performed for the 
equilibrium minima to compute the intrinsic level densities. 
The binding energy, pairing energy, entropy, and intrinsic level density as functions of nuclear 
temperature are displayed in Fig.~\ref{fig:Obs_oct}.
Just as in the case of the mass $A \approx 100$ region, the binding energies increase quadratically  with temperature, while the entropy first increase quadratically with $T$ below the critical temperature of pairing phase transition, and linearly for higher tempertaures. 
Fig.~\ref{fig:Obs_oct} (b) shows that the pairing collapse occurs at the critical temperature 
$T_{c} = 0.5 \sim 0.6$ MeV. The behaviour of energies, entropies, and intrinsic level densities as functions of $T$ is characterized by a discontinuity at $T_{c}$.

\begin{figure}[!]
 \includegraphics[width=0.48\textwidth]{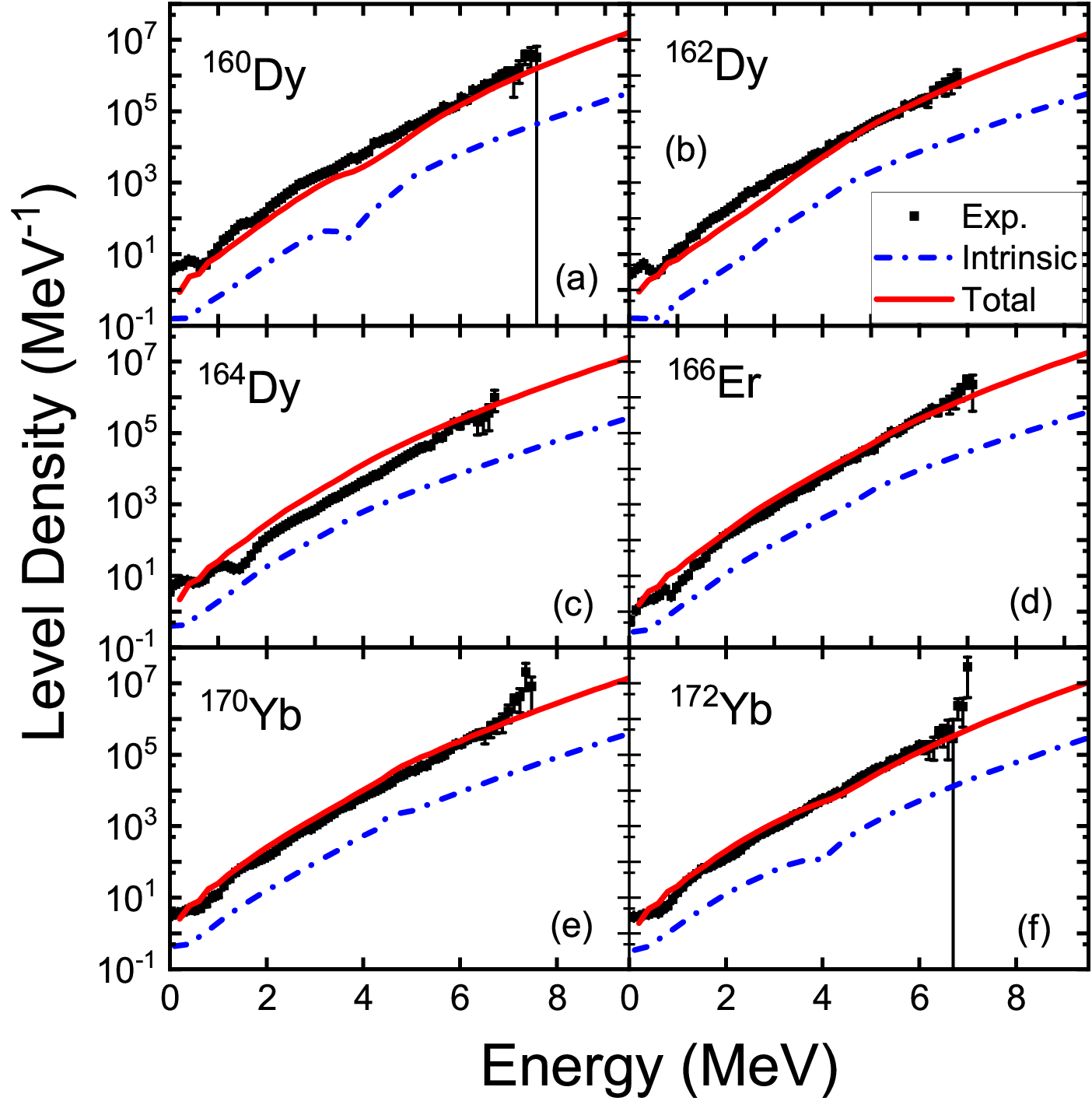}
\caption{(Color online)~\label{fig:LD_oct}%
The calculated intrinsic level densities (dash-dotted blue) and total level densities (solid red),  
as functions of excitation energy for $^{160,162,164}$Dy, $^{166}$Er, and $^{170,172}$Yb.
The data (black squares) are from 
Refs.~\cite{Guttormsen2003_PRC68-064306,Schiller2001_PRC63-021306,Nyhus2012_PRC85-014323,Melby2001_PRC63-044309,Agvaanluvsan2004_PRC70-054611}.
}
\end{figure}

\begin{figure}[!]
 \includegraphics[width=0.4\textwidth]{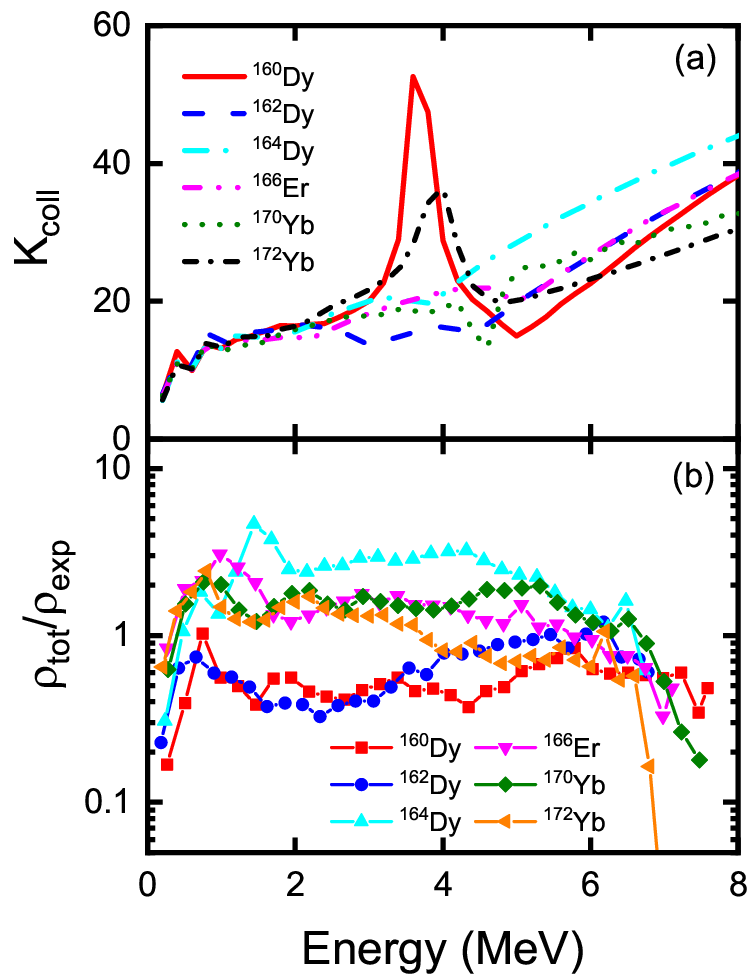}
\caption{(Color online)~\label{fig:Kcoll_Oct}%
Same as in the caption to Fig.~\ref{fig:Kcoll_tri} but for the axially-symmetric and reflection-asymmetric 
calculations of $^{160,162,164}$Dy, $^{166}$Er, and $^{170,172}$Yb.
}
\end{figure}

The self-consistent RHB energy surfaces and the corresponding ZPEs, the mass parameters, and the moments of inertial in the $(\beta_{2}, \beta_{3})$ plane at zero temperature, determine the 
axial quadrupole-octupole collective Hamiltonian Eq.~(\ref{eq:CHoct}). The eigenstates of this 
Hamiltonian are used to compute the total level densities (Eqs.~(\ref{eq:LDtot}) and (\ref{eq:LDcoll})). 
The calculated intrinsic level densities and the total level densities of 
$^{160,162,164}$Dy, $^{166}$Er, and $^{170,172}$Yb, as functions of the excitation energy, are compared in Fig.~\ref{fig:LD_oct} with the experimental values from 
Refs.~\cite{Guttormsen2003_PRC68-064306,Schiller2001_PRC63-021306,Nyhus2012_PRC85-014323,Melby2001_PRC63-044309,Agvaanluvsan2004_PRC70-054611}. Similar to the result obtained in the mass $A \approx 100$ region, the intrinsic level densities are systematically below the experimental values for all isotopes and all excitation energies. In the energy interval of measured values, the consistent microscopic calculation of the collective enhancement, using the axial quadrupole and octupole collective degrees of freedom relevant for this mass region, produces total level densities that are in very good agreement with available data. 

Figure \ref{fig:Kcoll_Oct} diplays the collective enhancement factors $K_{\rm{coll}}$ (a), and ratios $\rho_{\rm{tot}} / \rho_{\rm{exp}}$ of calculated and experimental level densities (b), as functions of excitation energy for $^{160,162,164}$Dy, $^{166}$Er, and $^{170,172}$Yb. Compared to the case of 
$A \approx 100$ nuclei, $K_{\rm{coll}}$ exhibits a more pronounced increase with energy. For $^{160}$Dy and  $^{172}$Yb one notices two strong peaks 
at $\approx 4$ MeV that correspond to the dips in the intrinsic level densities, caused by an unphysical collapse of pairing correlations in the FT-RHB calculation. Such discontinuities would be smoothed out by particle number projection, but this procedure is not included in the current version of the model. The overall agreement between theory and experiment is illustrated by the ratios $\rho_{\rm{tot}} / \rho_{\rm{exp}}$ in panel (b).  For $^{162,164}$Dy we can compare the values of $K_{\rm{coll}}$ at neutron separation energy with the recent prediction of the rotational enhancement factor $R_2$ averaged over angular momentum of Ref.~\cite{Grimes2019_PRC99-064331}. The predicted values of $R_2$: 45.3 at $S_n = 8.20$ Mev for $^{162}$Dy, and 
46.1 at $S_n = 7.63$ Mev for $^{164}$Dy (Table I of of Ref.~\cite{Grimes2019_PRC99-064331}), are very close to the corresponding collective 
enhancement factors obtained in the present microscopic calculation: $K_{\rm{coll}} = 39.8$ for $^{162}$Dy, and $K_{\rm{coll}} = 42.5$ for $^{164}$Dy. 

\section{\label{sec:summary}Summary}
A fully self-consistent microscopic approach for calculating nuclear level densities has been developed, based on global nuclear energy density functionals. The intrinsic level densities are computed in the thermodynamical approach using the saddle point approximation, with single-quasiparticle spectra obtained in a finite-temperature self-consistent mean-field (SCMF) calculation. In the present work we have used the finite-temperature relativistic Hartree-Bogoliubov (FT-RHB) model based on the DD-PC1 energy density functional and a finite-range pairing interaction. The total level densities are obtained by convoluting the intrinsic densities with the corresponding collective level densities. The collective levels are calculated as eigenstates of a five-dimensional quadrupole or quadrupole-octupole Hamiltonian, with parameters (mass parameters, moments of inertia, collective potential) fully determined by the SCMF calculation of the deformation energy surfaces and the corresponding single-quasiparticle levels as functions of the collective coordinates (shape variables). Therefore, in this approach both the intrinsic and collective level densities are completely determined by the choice of a global energy density functional and pairing interaction. One has to choose, however, the coordinates of the collective Hamiltonian depending on the specific nucleus under consideration. This is done for practical reasons, as the collective Hamiltonian can only take into account a small number of most relevant coordinates. For instance, quadrupole or quadrupole plus octupole shape variables will typically be used as collective coordinates. 

The model has been tested in several illustrative calculations in the $A \approx 100$ and $A \approx 160-170$ mass regions, where accurate experimental level densities are available in the energy interval  below the neutron separation energy. In the former region we have computed the level densities of 
$^{94,96,98}$Mo, $^{106,108}$Pd, and $^{106,112}$Cd. In general these nuclei exhibit equilibrium minima at moderate quadrupole deformation, and the deformation energy surfaces are rather soft in the $\gamma$ degree of freedom. Thus we have used the five-dimensional Hamiltonian in the quadrupole variables $\beta$ and $\gamma$ to calculate the levels that determine the collective enhancement of the intrinsic level densities. In the mass region of heavier nuclei level densities have been calculated for $^{160,162,164}$Dy, $^{166}$Er, and $^{170,172}$Yb. To a good approximation the equilibrium minima of these nuclei are axially quadrupole deformed, but also extended (soft) in the octupole deformation. In this case we have used an axially symmetric quadrupole-octupole Hamiltonian to calculate the collective level densities. 

In both mass regions it has been shown that, while the calculated intrinsic level densities reproduce the energy dependence of the data, their values are systematically too small and, therefore, additional degrees of freedom related to the shape of a nucleus have to be taken into account. The collective enhancement computed using the eigenstates of the five-dimensional quadrupole (mass $A \approx 100$) or axially symmetric quadrupole-octupole (mass  $A \approx 160-170$) Hamiltonian, yields total level densities that are in agreement with data in the entire interval of measured values. Since both the intrinsic and collective level densities are determined by the same underlying energy density functional and pairing interaction, the calculation is fully consistent and can be extended to any mass region and to nuclei far from stability for which data on collective levels are not available and, therefore, the semiempirical approaches to collective enhancement are not applicable. The method can be refined by improving the agreement of the collective levels with data and, of course, developed further by taking into account shape variables of higher multipolarity and/or the effect of pairing vibrations.  
\\
\bigskip
\acknowledgements
This work has been supported by the Inter-Governmental S\&T Cooperation Project between China and Croatia. 
It has also been supported in part by the QuantiXLie Centre of Excellence, a project co-financed by the Croatian Government and European Union through the European Regional Development Fund - the Competitiveness and Cohesion Operational Programme (KK.01.1.1.01)
and by the Croatian Science Foundation under the project Uncertainty quantification within the nuclear energy density 
framework (IP-2018-01-5987).
J.Z. acknowledges support by the National Natural Science Foundation of China under Grant No. 12005107 and No. 11790325.
Calculations have been performed in part at the HPC Cluster of KLTP/ITP-CAS and the Supercomputing Center,
Computer Network Information Center of CAS. 


\begin{thebibliography}{66}%
\makeatletter
\providecommand \@ifxundefined [1]{%
 \@ifx{#1\undefined}
}%
\providecommand \@ifnum [1]{%
 \ifnum #1\expandafter \@firstoftwo
 \else \expandafter \@secondoftwo
 \fi
}%
\providecommand \@ifx [1]{%
 \ifx #1\expandafter \@firstoftwo
 \else \expandafter \@secondoftwo
 \fi
}%
\providecommand \natexlab [1]{#1}%
\providecommand \enquote  [1]{``#1''}%
\providecommand \bibnamefont  [1]{#1}%
\providecommand \bibfnamefont [1]{#1}%
\providecommand \citenamefont [1]{#1}%
\providecommand \href@noop [0]{\@secondoftwo}%
\providecommand \href [0]{\begingroup \@sanitize@url \@href}%
\providecommand \@href[1]{\@@startlink{#1}\@@href}%
\providecommand \@@href[1]{\endgroup#1\@@endlink}%
\providecommand \@sanitize@url [0]{\catcode `\\12\catcode `\$12\catcode
  `\&12\catcode `\#12\catcode `\^12\catcode `\_12\catcode `\%12\relax}%
\providecommand \@@startlink[1]{}%
\providecommand \@@endlink[0]{}%
\providecommand \url  [0]{\begingroup\@sanitize@url \@url }%
\providecommand \@url [1]{\endgroup\@href {#1}{\urlprefix }}%
\providecommand \urlprefix  [0]{URL }%
\providecommand \Eprint [0]{\href }%
\providecommand \doibase [0]{http://dx.doi.org/}%
\providecommand \selectlanguage [0]{\@gobble}%
\providecommand \bibinfo  [0]{\@secondoftwo}%
\providecommand \bibfield  [0]{\@secondoftwo}%
\providecommand \translation [1]{[#1]}%
\providecommand \BibitemOpen [0]{}%
\providecommand \bibitemStop [0]{}%
\providecommand \bibitemNoStop [0]{.\EOS\space}%
\providecommand \EOS [0]{\spacefactor3000\relax}%
\providecommand \BibitemShut  [1]{\csname bibitem#1\endcsname}%
\let\auto@bib@innerbib\@empty
\bibitem [{\citenamefont {Bethe}(1937)}]{Bethe1937_RMP9-69}%
  \BibitemOpen
  \bibfield  {author} {\bibinfo {author} {\bibfnamefont {H.~A.}\ \bibnamefont
  {Bethe}},\ }\href {\doibase 10.1103/revmodphys.9.69} {\bibfield  {journal}
  {\bibinfo  {journal} {Rev. Mod. Phys.}\ }\textbf {\bibinfo {volume} {9}},\
  \bibinfo {pages} {69} (\bibinfo {year} {1937})}\BibitemShut {NoStop}%
\bibitem [{\citenamefont {Koning}\ \emph {et~al.}(2008)\citenamefont {Koning},
  \citenamefont {Hilaire},\ and\ \citenamefont
  {Goriely}}]{Koning2008_NPA810-13}%
  \BibitemOpen
  \bibfield  {author} {\bibinfo {author} {\bibfnamefont {A.}~\bibnamefont
  {Koning}}, \bibinfo {author} {\bibfnamefont {S.}~\bibnamefont {Hilaire}}, \
  and\ \bibinfo {author} {\bibfnamefont {S.}~\bibnamefont {Goriely}},\ }\href
  {\doibase 10.1016/j.nuclphysa.2008.06.005} {\bibfield  {journal} {\bibinfo
  {journal} {Nucl. Phys. A}\ }\textbf {\bibinfo {volume} {810}},\ \bibinfo
  {pages} {13} (\bibinfo {year} {2008})}\BibitemShut {NoStop}%
\bibitem [{\citenamefont {Goriely}(1996)}]{Goriely1996_NPA605-28}%
  \BibitemOpen
  \bibfield  {author} {\bibinfo {author} {\bibfnamefont {S.}~\bibnamefont
  {Goriely}},\ }\href {\doibase 10.1016/0375-9474(96)00162-5} {\bibfield
  {journal} {\bibinfo  {journal} {Nucl. Phys. A}\ }\textbf {\bibinfo {volume}
  {605}},\ \bibinfo {pages} {28} (\bibinfo {year} {1996})}\BibitemShut
  {NoStop}%
\bibitem [{\citenamefont {Gross}\ and\ \citenamefont
  {Heck}(1993)}]{Gross1993_PLB318-405}%
  \BibitemOpen
  \bibfield  {author} {\bibinfo {author} {\bibfnamefont {D.}~\bibnamefont
  {Gross}}\ and\ \bibinfo {author} {\bibfnamefont {R.}~\bibnamefont {Heck}},\
  }\href {\doibase 10.1016/0370-2693(93)91531-q} {\bibfield  {journal}
  {\bibinfo  {journal} {Phys. Lett. B}\ }\textbf {\bibinfo {volume} {318}},\
  \bibinfo {pages} {405} (\bibinfo {year} {1993})}\BibitemShut {NoStop}%
\bibitem [{\citenamefont {Alhassid}\ \emph {et~al.}(1999)\citenamefont
  {Alhassid}, \citenamefont {Liu},\ and\ \citenamefont
  {Nakada}}]{Alhassid1999_PRL83-4265}%
  \BibitemOpen
  \bibfield  {author} {\bibinfo {author} {\bibfnamefont {Y.}~\bibnamefont
  {Alhassid}}, \bibinfo {author} {\bibfnamefont {S.}~\bibnamefont {Liu}}, \
  and\ \bibinfo {author} {\bibfnamefont {H.}~\bibnamefont {Nakada}},\ }\href
  {\doibase 10.1103/physrevlett.83.4265} {\bibfield  {journal} {\bibinfo
  {journal} {Phys. Rev. Lett.}\ }\textbf {\bibinfo {volume} {83}},\ \bibinfo
  {pages} {4265} (\bibinfo {year} {1999})}\BibitemShut {NoStop}%
\bibitem [{\citenamefont {Alhassid}\ \emph {et~al.}(2007)\citenamefont
  {Alhassid}, \citenamefont {Liu},\ and\ \citenamefont
  {Nakada}}]{Alhassid2007_PRL99-162504}%
  \BibitemOpen
  \bibfield  {author} {\bibinfo {author} {\bibfnamefont {Y.}~\bibnamefont
  {Alhassid}}, \bibinfo {author} {\bibfnamefont {S.}~\bibnamefont {Liu}}, \
  and\ \bibinfo {author} {\bibfnamefont {H.}~\bibnamefont {Nakada}},\ }\href
  {\doibase 10.1103/physrevlett.99.162504} {\bibfield  {journal} {\bibinfo
  {journal} {Phys. Rev. Lett.}\ }\textbf {\bibinfo {volume} {99}},\ \bibinfo
  {pages} {162504} (\bibinfo {year} {2007})}\BibitemShut {NoStop}%
\bibitem [{\citenamefont {Alhassid}\ \emph {et~al.}(2015)\citenamefont
  {Alhassid}, \citenamefont {Bonett-Matiz}, \citenamefont {Liu},\ and\
  \citenamefont {Nakada}}]{Alhassid2015_PRC92-024307}%
  \BibitemOpen
  \bibfield  {author} {\bibinfo {author} {\bibfnamefont {Y.}~\bibnamefont
  {Alhassid}}, \bibinfo {author} {\bibfnamefont {M.}~\bibnamefont
  {Bonett-Matiz}}, \bibinfo {author} {\bibfnamefont {S.}~\bibnamefont {Liu}}, \
  and\ \bibinfo {author} {\bibfnamefont {H.}~\bibnamefont {Nakada}},\ }\href
  {\doibase 10.1103/physrevc.92.024307} {\bibfield  {journal} {\bibinfo
  {journal} {Phys. Rev. C}\ }\textbf {\bibinfo {volume} {92}},\ \bibinfo
  {pages} {024307} (\bibinfo {year} {2015})}\BibitemShut {NoStop}%
\bibitem [{\citenamefont {Zelevinsky}\ \emph {et~al.}(2018)\citenamefont
  {Zelevinsky}, \citenamefont {Karampagia},\ and\ \citenamefont
  {Berlaga}}]{Zelevinsky2018_PLB783-428}%
  \BibitemOpen
  \bibfield  {author} {\bibinfo {author} {\bibfnamefont {V.}~\bibnamefont
  {Zelevinsky}}, \bibinfo {author} {\bibfnamefont {S.}~\bibnamefont
  {Karampagia}}, \ and\ \bibinfo {author} {\bibfnamefont {A.}~\bibnamefont
  {Berlaga}},\ }\href
  {http://www.sciencedirect.com/science/article/pii/S0370269318305616}
  {\bibfield  {journal} {\bibinfo  {journal} {Phys. Lett. B}\ }\textbf
  {\bibinfo {volume} {783}},\ \bibinfo {pages} {428} (\bibinfo {year}
  {2018})}\BibitemShut {NoStop}%
\bibitem [{\citenamefont {Sen'kov}\ and\ \citenamefont
  {Horoi}(2010)}]{Senkov2010_PRC82-024304}%
  \BibitemOpen
  \bibfield  {author} {\bibinfo {author} {\bibfnamefont {R.~A.}\ \bibnamefont
  {Sen'kov}}\ and\ \bibinfo {author} {\bibfnamefont {M.}~\bibnamefont
  {Horoi}},\ }\href {\doibase 10.1103/physrevc.82.024304} {\bibfield  {journal}
  {\bibinfo  {journal} {Phys. Rev. C}\ }\textbf {\bibinfo {volume} {82}},\
  \bibinfo {pages} {024304} (\bibinfo {year} {2010})}\BibitemShut {NoStop}%
\bibitem [{\citenamefont {Shimizu}\ \emph {et~al.}(2016)\citenamefont
  {Shimizu}, \citenamefont {Utsuno}, \citenamefont {Futamura}, \citenamefont
  {Sakurai}, \citenamefont {Mizusaki},\ and\ \citenamefont
  {Otsuka}}]{Shimizu2016_PLB753-13}%
  \BibitemOpen
  \bibfield  {author} {\bibinfo {author} {\bibfnamefont {N.}~\bibnamefont
  {Shimizu}}, \bibinfo {author} {\bibfnamefont {Y.}~\bibnamefont {Utsuno}},
  \bibinfo {author} {\bibfnamefont {Y.}~\bibnamefont {Futamura}}, \bibinfo
  {author} {\bibfnamefont {T.}~\bibnamefont {Sakurai}}, \bibinfo {author}
  {\bibfnamefont {T.}~\bibnamefont {Mizusaki}}, \ and\ \bibinfo {author}
  {\bibfnamefont {T.}~\bibnamefont {Otsuka}},\ }\href {\doibase
  10.1016/j.physletb.2015.12.005} {\bibfield  {journal} {\bibinfo  {journal}
  {Phys. Lett. B}\ }\textbf {\bibinfo {volume} {753}},\ \bibinfo {pages} {13}
  (\bibinfo {year} {2016})}\BibitemShut {NoStop}%
\bibitem [{\citenamefont {Ormand}\ and\ \citenamefont
  {Brown}(2020)}]{Ormand2020_PRC102-014315}%
  \BibitemOpen
  \bibfield  {author} {\bibinfo {author} {\bibfnamefont {W.~E.}\ \bibnamefont
  {Ormand}}\ and\ \bibinfo {author} {\bibfnamefont {B.~A.}\ \bibnamefont
  {Brown}},\ }\href {\doibase 10.1103/physrevc.102.014315} {\bibfield
  {journal} {\bibinfo  {journal} {Phys. Rev. C}\ }\textbf {\bibinfo {volume}
  {102}},\ \bibinfo {pages} {014315} (\bibinfo {year} {2020})}\BibitemShut
  {NoStop}%
\bibitem [{\citenamefont {Kolomietz}\ \emph {et~al.}(2018)\citenamefont
  {Kolomietz}, \citenamefont {Sanzhur},\ and\ \citenamefont
  {Shlomo}}]{Kolomietz2018_PRC97-064302}%
  \BibitemOpen
  \bibfield  {author} {\bibinfo {author} {\bibfnamefont {V.~M.}\ \bibnamefont
  {Kolomietz}}, \bibinfo {author} {\bibfnamefont {A.~I.}\ \bibnamefont
  {Sanzhur}}, \ and\ \bibinfo {author} {\bibfnamefont {S.}~\bibnamefont
  {Shlomo}},\ }\href {https://link.aps.org/doi/10.1103/PhysRevC.97.064302}
  {\bibfield  {journal} {\bibinfo  {journal} {Phys. Rev. C}\ }\textbf {\bibinfo
  {volume} {97}},\ \bibinfo {pages} {064302} (\bibinfo {year}
  {2018})}\BibitemShut {NoStop}%
\bibitem [{\citenamefont {Hung}\ \emph {et~al.}(2017)\citenamefont {Hung},
  \citenamefont {Dang},\ and\ \citenamefont {Huong}}]{Hung2017_PRL118-022502}%
  \BibitemOpen
  \bibfield  {author} {\bibinfo {author} {\bibfnamefont {N.~Q.}\ \bibnamefont
  {Hung}}, \bibinfo {author} {\bibfnamefont {N.~D.}\ \bibnamefont {Dang}}, \
  and\ \bibinfo {author} {\bibfnamefont {L.~T.~Q.}\ \bibnamefont {Huong}},\
  }\href {http://link.aps.org/doi/10.1103/PhysRevLett.118.022502} {\bibfield
  {journal} {\bibinfo  {journal} {Phys. Rev. Lett.}\ }\textbf {\bibinfo
  {volume} {118}},\ \bibinfo {pages} {022502} (\bibinfo {year}
  {2017})}\BibitemShut {NoStop}%
\bibitem [{\citenamefont {Dang}\ \emph {et~al.}(2017)\citenamefont {Dang},
  \citenamefont {Hung},\ and\ \citenamefont {Huong}}]{Dang2017_PRC96-054321}%
  \BibitemOpen
  \bibfield  {author} {\bibinfo {author} {\bibfnamefont {N.~D.}\ \bibnamefont
  {Dang}}, \bibinfo {author} {\bibfnamefont {N.~Q.}\ \bibnamefont {Hung}}, \
  and\ \bibinfo {author} {\bibfnamefont {L.~T.~Q.}\ \bibnamefont {Huong}},\
  }\href {\doibase 10.1103/physrevc.96.054321} {\bibfield  {journal} {\bibinfo
  {journal} {Phys. Rev. C}\ }\textbf {\bibinfo {volume} {96}},\ \bibinfo
  {pages} {054321} (\bibinfo {year} {2017})}\BibitemShut {NoStop}%
\bibitem [{\citenamefont {Dey}\ \emph {et~al.}(2017)\citenamefont {Dey},
  \citenamefont {Pandit}, \citenamefont {Bhattacharya}, \citenamefont {Hung},
  \citenamefont {Dang}, \citenamefont {Phuc}, \citenamefont {Mondal},
  \citenamefont {Mukhopadhyay}, \citenamefont {Pal}, \citenamefont {De},\ and\
  \citenamefont {Banerjee}}]{Dey2017_PRC96-054326}%
  \BibitemOpen
  \bibfield  {author} {\bibinfo {author} {\bibfnamefont {B.}~\bibnamefont
  {Dey}}, \bibinfo {author} {\bibfnamefont {D.}~\bibnamefont {Pandit}},
  \bibinfo {author} {\bibfnamefont {S.}~\bibnamefont {Bhattacharya}}, \bibinfo
  {author} {\bibfnamefont {N.~Q.}\ \bibnamefont {Hung}}, \bibinfo {author}
  {\bibfnamefont {N.~D.}\ \bibnamefont {Dang}}, \bibinfo {author}
  {\bibfnamefont {L.~T.}\ \bibnamefont {Phuc}}, \bibinfo {author}
  {\bibfnamefont {D.}~\bibnamefont {Mondal}}, \bibinfo {author} {\bibfnamefont
  {S.}~\bibnamefont {Mukhopadhyay}}, \bibinfo {author} {\bibfnamefont
  {S.}~\bibnamefont {Pal}}, \bibinfo {author} {\bibfnamefont {A.}~\bibnamefont
  {De}}, \ and\ \bibinfo {author} {\bibfnamefont {S.~R.}\ \bibnamefont
  {Banerjee}},\ }\href {\doibase 10.1103/physrevc.96.054326} {\bibfield
  {journal} {\bibinfo  {journal} {Phys. Rev. C}\ }\textbf {\bibinfo {volume}
  {96}},\ \bibinfo {pages} {054326} (\bibinfo {year} {2017})}\BibitemShut
  {NoStop}%
\bibitem [{\citenamefont {Dey}\ \emph {et~al.}(2019)\citenamefont {Dey},
  \citenamefont {Hung}, \citenamefont {Pandit}, \citenamefont {Bhattacharya},
  \citenamefont {Dang}, \citenamefont {Huong}, \citenamefont {Mondal},
  \citenamefont {Mukhopadhyay}, \citenamefont {Pal}, \citenamefont {De},\ and\
  \citenamefont {Banerjee}}]{Dey2019_PLB789-634}%
  \BibitemOpen
  \bibfield  {author} {\bibinfo {author} {\bibfnamefont {B.}~\bibnamefont
  {Dey}}, \bibinfo {author} {\bibfnamefont {N.~Q.}\ \bibnamefont {Hung}},
  \bibinfo {author} {\bibfnamefont {D.}~\bibnamefont {Pandit}}, \bibinfo
  {author} {\bibfnamefont {S.}~\bibnamefont {Bhattacharya}}, \bibinfo {author}
  {\bibfnamefont {N.~D.}\ \bibnamefont {Dang}}, \bibinfo {author}
  {\bibfnamefont {L.~Q.}\ \bibnamefont {Huong}}, \bibinfo {author}
  {\bibfnamefont {D.}~\bibnamefont {Mondal}}, \bibinfo {author} {\bibfnamefont
  {S.}~\bibnamefont {Mukhopadhyay}}, \bibinfo {author} {\bibfnamefont
  {S.}~\bibnamefont {Pal}}, \bibinfo {author} {\bibfnamefont {A.}~\bibnamefont
  {De}}, \ and\ \bibinfo {author} {\bibfnamefont {S.}~\bibnamefont
  {Banerjee}},\ }\href {\doibase 10.1016/j.physletb.2018.12.007} {\bibfield
  {journal} {\bibinfo  {journal} {Phys. Lett. B}\ }\textbf {\bibinfo {volume}
  {789}},\ \bibinfo {pages} {634} (\bibinfo {year} {2019})}\BibitemShut
  {NoStop}%
\bibitem [{\citenamefont {Choudhury}\ and\ \citenamefont
  {Gupta}(1977)}]{Choudhury1977_PRC16-757}%
  \BibitemOpen
  \bibfield  {author} {\bibinfo {author} {\bibfnamefont {F.~N.}\ \bibnamefont
  {Choudhury}}\ and\ \bibinfo {author} {\bibfnamefont {S.~D.}\ \bibnamefont
  {Gupta}},\ }\href {https://link.aps.org/doi/10.1103/PhysRevC.16.757}
  {\bibfield  {journal} {\bibinfo  {journal} {Phys. Rev. C}\ }\textbf {\bibinfo
  {volume} {16}},\ \bibinfo {pages} {757} (\bibinfo {year} {1977})}\BibitemShut
  {NoStop}%
\bibitem [{\citenamefont {Minato}(2011)}]{Minato2011_JNST48-984}%
  \BibitemOpen
  \bibfield  {author} {\bibinfo {author} {\bibfnamefont {F.}~\bibnamefont
  {Minato}},\ }\href {\doibase 10.1080/18811248.2011.9711785} {\bibfield
  {journal} {\bibinfo  {journal} {J. Nucl. Sci. Technol.}\ }\textbf {\bibinfo
  {volume} {48}},\ \bibinfo {pages} {984} (\bibinfo {year} {2011})}\BibitemShut
  {NoStop}%
\bibitem [{\citenamefont {Demetriou}\ and\ \citenamefont
  {Goriely}(2001)}]{Demetriou2001_NPA695-95}%
  \BibitemOpen
  \bibfield  {author} {\bibinfo {author} {\bibfnamefont {P.}~\bibnamefont
  {Demetriou}}\ and\ \bibinfo {author} {\bibfnamefont {S.}~\bibnamefont
  {Goriely}},\ }\href
  {http://www.sciencedirect.com/science/article/pii/S0375947401010958}
  {\bibfield  {journal} {\bibinfo  {journal} {Nucl. Phys. A}\ }\textbf
  {\bibinfo {volume} {695}},\ \bibinfo {pages} {95} (\bibinfo {year}
  {2001})}\BibitemShut {NoStop}%
\bibitem [{\citenamefont {Hilaire}\ and\ \citenamefont
  {Goriely}(2006)}]{Hilaire2006_NPA779-63}%
  \BibitemOpen
  \bibfield  {author} {\bibinfo {author} {\bibfnamefont {S.}~\bibnamefont
  {Hilaire}}\ and\ \bibinfo {author} {\bibfnamefont {S.}~\bibnamefont
  {Goriely}},\ }\href
  {http://www.sciencedirect.com/science/article/pii/S037594740600580X}
  {\bibfield  {journal} {\bibinfo  {journal} {Nucl. Phys. A}\ }\textbf
  {\bibinfo {volume} {779}},\ \bibinfo {pages} {63} (\bibinfo {year}
  {2006})}\BibitemShut {NoStop}%
\bibitem [{\citenamefont {Hilaire}\ \emph {et~al.}(2001)\citenamefont
  {Hilaire}, \citenamefont {Delaroche},\ and\ \citenamefont
  {Girod}}]{Hilaire2001_TEPJA12-169}%
  \BibitemOpen
  \bibfield  {author} {\bibinfo {author} {\bibfnamefont {S.}~\bibnamefont
  {Hilaire}}, \bibinfo {author} {\bibfnamefont {J.~P.}\ \bibnamefont
  {Delaroche}}, \ and\ \bibinfo {author} {\bibfnamefont {M.}~\bibnamefont
  {Girod}},\ }\href {https://doi.org/10.1007/s100500170025} {\bibfield
  {journal} {\bibinfo  {journal} {Eur. Phys. J. A}\ }\textbf {\bibinfo {volume}
  {12}},\ \bibinfo {pages} {169} (\bibinfo {year} {2001})}\BibitemShut
  {NoStop}%
\bibitem [{\citenamefont {Goriely}\ \emph {et~al.}(2008)\citenamefont
  {Goriely}, \citenamefont {Hilaire},\ and\ \citenamefont
  {Koning}}]{Goriely2008_PRC78-064307}%
  \BibitemOpen
  \bibfield  {author} {\bibinfo {author} {\bibfnamefont {S.}~\bibnamefont
  {Goriely}}, \bibinfo {author} {\bibfnamefont {S.}~\bibnamefont {Hilaire}}, \
  and\ \bibinfo {author} {\bibfnamefont {A.~J.}\ \bibnamefont {Koning}},\
  }\href {https://link.aps.org/doi/10.1103/PhysRevC.78.064307} {\bibfield
  {journal} {\bibinfo  {journal} {Phys. Rev. C}\ }\textbf {\bibinfo {volume}
  {78}},\ \bibinfo {pages} {064307} (\bibinfo {year} {2008})}\BibitemShut
  {NoStop}%
\bibitem [{\citenamefont {Bohr}\ and\ \citenamefont
  {Mottelson}(1969)}]{Bohr1969_Nucl_Structure}%
  \BibitemOpen
  \bibfield  {author} {\bibinfo {author} {\bibfnamefont {A.}~\bibnamefont
  {Bohr}}\ and\ \bibinfo {author} {\bibfnamefont {B.~R.}\ \bibnamefont
  {Mottelson}},\ }\href@noop {} {\emph {\bibinfo {title} {Nuclear
  Structure}}},\ \bibinfo {edition} {1st}\ ed.,\ Vol.~\bibinfo {volume} {I}\
  (\bibinfo  {publisher} {Benjamin, New York},\ \bibinfo {year}
  {1969})\BibitemShut {NoStop}%
\bibitem [{\citenamefont {Junghans}\ \emph {et~al.}(1998)\citenamefont
  {Junghans}, \citenamefont {de~Jong}, \citenamefont {Clerc}, \citenamefont
  {Ignatyuk}, \citenamefont {Kudyaev},\ and\ \citenamefont
  {Schmidt}}]{Junghans1998_NPA629-635}%
  \BibitemOpen
  \bibfield  {author} {\bibinfo {author} {\bibfnamefont {A.}~\bibnamefont
  {Junghans}}, \bibinfo {author} {\bibfnamefont {M.}~\bibnamefont {de~Jong}},
  \bibinfo {author} {\bibfnamefont {H.-G.}\ \bibnamefont {Clerc}}, \bibinfo
  {author} {\bibfnamefont {A.}~\bibnamefont {Ignatyuk}}, \bibinfo {author}
  {\bibfnamefont {G.}~\bibnamefont {Kudyaev}}, \ and\ \bibinfo {author}
  {\bibfnamefont {K.-H.}\ \bibnamefont {Schmidt}},\ }\href {\doibase
  10.1016/s0375-9474(98)00658-7} {\bibfield  {journal} {\bibinfo  {journal}
  {Nucl. Phys. A}\ }\textbf {\bibinfo {volume} {629}},\ \bibinfo {pages} {635}
  (\bibinfo {year} {1998})}\BibitemShut {NoStop}%
\bibitem [{\citenamefont {Rahmatinejad}\ \emph {et~al.}(2020)\citenamefont
  {Rahmatinejad}, \citenamefont {Shneidman}, \citenamefont {Antonenko},
  \citenamefont {Bezbakh}, \citenamefont {Adamian},\ and\ \citenamefont
  {Malov}}]{Rahmatinejad2020_PRC101-054315}%
  \BibitemOpen
  \bibfield  {author} {\bibinfo {author} {\bibfnamefont {A.}~\bibnamefont
  {Rahmatinejad}}, \bibinfo {author} {\bibfnamefont {T.~M.}\ \bibnamefont
  {Shneidman}}, \bibinfo {author} {\bibfnamefont {N.~V.}\ \bibnamefont
  {Antonenko}}, \bibinfo {author} {\bibfnamefont {A.~N.}\ \bibnamefont
  {Bezbakh}}, \bibinfo {author} {\bibfnamefont {G.~G.}\ \bibnamefont
  {Adamian}}, \ and\ \bibinfo {author} {\bibfnamefont {L.~A.}\ \bibnamefont
  {Malov}},\ }\href {\doibase 10.1103/physrevc.101.054315} {\bibfield
  {journal} {\bibinfo  {journal} {Phys. Rev. C}\ }\textbf {\bibinfo {volume}
  {101}},\ \bibinfo {pages} {054315} (\bibinfo {year} {2020})}\BibitemShut
  {NoStop}%
\bibitem [{\citenamefont {Kargar}(2007)}]{Kargar2007_PRC75-064319}%
  \BibitemOpen
  \bibfield  {author} {\bibinfo {author} {\bibfnamefont {Z.}~\bibnamefont
  {Kargar}},\ }\href {\doibase 10.1103/physrevc.75.064319} {\bibfield
  {journal} {\bibinfo  {journal} {Phys. Rev. C}\ }\textbf {\bibinfo {volume}
  {75}},\ \bibinfo {pages} {064319} (\bibinfo {year} {2007})}\BibitemShut
  {NoStop}%
\bibitem [{\citenamefont {Grimes}\ \emph {et~al.}(2019)\citenamefont {Grimes},
  \citenamefont {Massey},\ and\ \citenamefont
  {Voinov}}]{Grimes2019_PRC99-064331}%
  \BibitemOpen
  \bibfield  {author} {\bibinfo {author} {\bibfnamefont {S.~M.}\ \bibnamefont
  {Grimes}}, \bibinfo {author} {\bibfnamefont {T.~N.}\ \bibnamefont {Massey}},
  \ and\ \bibinfo {author} {\bibfnamefont {A.~V.}\ \bibnamefont {Voinov}},\
  }\href {\doibase 10.1103/physrevc.99.064331} {\bibfield  {journal} {\bibinfo
  {journal} {Phys. Rev. C}\ }\textbf {\bibinfo {volume} {99}},\ \bibinfo
  {pages} {064331} (\bibinfo {year} {2019})}\BibitemShut {NoStop}%
\bibitem [{\citenamefont {Goriely}\ \emph {et~al.}(2009)\citenamefont
  {Goriely}, \citenamefont {Hilaire}, \citenamefont {Koning}, \citenamefont
  {Sin},\ and\ \citenamefont {Capote}}]{Goriely2009_PRC79-024612}%
  \BibitemOpen
  \bibfield  {author} {\bibinfo {author} {\bibfnamefont {S.}~\bibnamefont
  {Goriely}}, \bibinfo {author} {\bibfnamefont {S.}~\bibnamefont {Hilaire}},
  \bibinfo {author} {\bibfnamefont {A.~J.}\ \bibnamefont {Koning}}, \bibinfo
  {author} {\bibfnamefont {M.}~\bibnamefont {Sin}}, \ and\ \bibinfo {author}
  {\bibfnamefont {R.}~\bibnamefont {Capote}},\ }\href
  {https://link.aps.org/doi/10.1103/PhysRevC.79.024612} {\bibfield  {journal}
  {\bibinfo  {journal} {Phys. Rev. C}\ }\textbf {\bibinfo {volume} {79}},\
  \bibinfo {pages} {024612} (\bibinfo {year} {2009})}\BibitemShut {NoStop}%
\bibitem [{\citenamefont {Goriely}\ \emph {et~al.}(2011)\citenamefont
  {Goriely}, \citenamefont {Hilaire}, \citenamefont {Koning},\ and\
  \citenamefont {Capote}}]{Goriely2011_PRC83-034601}%
  \BibitemOpen
  \bibfield  {author} {\bibinfo {author} {\bibfnamefont {S.}~\bibnamefont
  {Goriely}}, \bibinfo {author} {\bibfnamefont {S.}~\bibnamefont {Hilaire}},
  \bibinfo {author} {\bibfnamefont {A.~J.}\ \bibnamefont {Koning}}, \ and\
  \bibinfo {author} {\bibfnamefont {R.}~\bibnamefont {Capote}},\ }\href
  {https://link.aps.org/doi/10.1103/PhysRevC.83.034601} {\bibfield  {journal}
  {\bibinfo  {journal} {Phys. Rev. C}\ }\textbf {\bibinfo {volume} {83}},\
  \bibinfo {pages} {034601} (\bibinfo {year} {2011})}\BibitemShut {NoStop}%
\bibitem [{\citenamefont {Ward}\ \emph {et~al.}(2017)\citenamefont {Ward},
  \citenamefont {Carlsson}, \citenamefont {D{\o}ssing}, \citenamefont
  {M{\"o}ller}, \citenamefont {Randrup},\ and\ \citenamefont
  {{\AA}berg}}]{Ward2017_PRC95-024618}%
  \BibitemOpen
  \bibfield  {author} {\bibinfo {author} {\bibfnamefont {D.~E.}\ \bibnamefont
  {Ward}}, \bibinfo {author} {\bibfnamefont {B.~G.}\ \bibnamefont {Carlsson}},
  \bibinfo {author} {\bibfnamefont {T.}~\bibnamefont {D{\o}ssing}}, \bibinfo
  {author} {\bibfnamefont {P.}~\bibnamefont {M{\"o}ller}}, \bibinfo {author}
  {\bibfnamefont {J.}~\bibnamefont {Randrup}}, \ and\ \bibinfo {author}
  {\bibfnamefont {S.}~\bibnamefont {{\AA}berg}},\ }\href
  {http://link.aps.org/doi/10.1103/PhysRevC.95.024618} {\bibfield  {journal}
  {\bibinfo  {journal} {Phys. Rev. C}\ }\textbf {\bibinfo {volume} {95}},\
  \bibinfo {pages} {024618} (\bibinfo {year} {2017})}\BibitemShut {NoStop}%
\bibitem [{\citenamefont {Vretenar}\ \emph {et~al.}(2005)\citenamefont
  {Vretenar}, \citenamefont {Afanasjev}, \citenamefont {Lalazissis},\ and\
  \citenamefont {Ring}}]{Vretenar2005_PR409-101}%
  \BibitemOpen
  \bibfield  {author} {\bibinfo {author} {\bibfnamefont {D.}~\bibnamefont
  {Vretenar}}, \bibinfo {author} {\bibfnamefont {A.}~\bibnamefont {Afanasjev}},
  \bibinfo {author} {\bibfnamefont {G.}~\bibnamefont {Lalazissis}}, \ and\
  \bibinfo {author} {\bibfnamefont {P.}~\bibnamefont {Ring}},\ }\href {\doibase
  DOI: 10.1016/j.physrep.2004.10.001} {\bibfield  {journal} {\bibinfo
  {journal} {Phys. Rep.}\ }\textbf {\bibinfo {volume} {409}},\ \bibinfo {pages}
  {101 } (\bibinfo {year} {2005})}\BibitemShut {NoStop}%
\bibitem [{\citenamefont {Meng}\ \emph {et~al.}(2006)\citenamefont {Meng},
  \citenamefont {Toki}, \citenamefont {Zhou}, \citenamefont {Zhang},
  \citenamefont {Long},\ and\ \citenamefont {Geng}}]{Meng2006_PPNP57-470}%
  \BibitemOpen
  \bibfield  {author} {\bibinfo {author} {\bibfnamefont {J.}~\bibnamefont
  {Meng}}, \bibinfo {author} {\bibfnamefont {H.}~\bibnamefont {Toki}}, \bibinfo
  {author} {\bibfnamefont {S.}~\bibnamefont {Zhou}}, \bibinfo {author}
  {\bibfnamefont {S.}~\bibnamefont {Zhang}}, \bibinfo {author} {\bibfnamefont
  {W.}~\bibnamefont {Long}}, \ and\ \bibinfo {author} {\bibfnamefont
  {L.}~\bibnamefont {Geng}},\ }\href {\doibase DOI: 10.1016/j.ppnp.2005.06.001}
  {\bibfield  {journal} {\bibinfo  {journal} {Prog. Part. Nucl. Phys.}\
  }\textbf {\bibinfo {volume} {57}},\ \bibinfo {pages} {470 } (\bibinfo {year}
  {2006})}\BibitemShut {NoStop}%
\bibitem [{\citenamefont {Zhou}(2016)}]{Zhou2016_PS91-063008}%
  \BibitemOpen
  \bibfield  {author} {\bibinfo {author} {\bibfnamefont {S.-G.}\ \bibnamefont
  {Zhou}},\ }\href {\doibase 10.1088/0031-8949/91/6/063008} {\bibfield
  {journal} {\bibinfo  {journal} {Phys. Scr.}\ }\textbf {\bibinfo {volume}
  {91}},\ \bibinfo {pages} {063008} (\bibinfo {year} {2016})}\BibitemShut
  {NoStop}%
\bibitem [{\citenamefont {Meng}(2016)}]{Meng2016_WorldSci}%
  \BibitemOpen
  \bibinfo {editor} {\bibfnamefont {J.}~\bibnamefont {Meng}},\ ed.,\ \href@noop
  {} {\emph {\bibinfo {title} {Relativistic Density Functional for Nuclear
  Structure}}},\ \bibinfo {series} {International Review of Nuclear Physics},
  Vol.~\bibinfo {volume} {10}\ (\bibinfo  {publisher} {World Scientific},\
  \bibinfo {year} {2016})\BibitemShut {NoStop}%
\bibitem [{\citenamefont {Li}\ \emph {et~al.}(2016)\citenamefont {Li},
  \citenamefont {Niksic},\ and\ \citenamefont
  {Vretenar}}]{Li2016_JPG43-024005}%
  \BibitemOpen
  \bibfield  {author} {\bibinfo {author} {\bibfnamefont {Z.~P.}\ \bibnamefont
  {Li}}, \bibinfo {author} {\bibfnamefont {T.}~\bibnamefont {Niksic}}, \ and\
  \bibinfo {author} {\bibfnamefont {D.}~\bibnamefont {Vretenar}},\ }\href
  {http://stacks.iop.org/0954-3899/43/i=2/a=024005} {\bibfield  {journal}
  {\bibinfo  {journal} {J. Phys. G: Nucl. Part. Phys.}\ }\textbf {\bibinfo
  {volume} {43}},\ \bibinfo {pages} {024005} (\bibinfo {year}
  {2016})}\BibitemShut {NoStop}%
\bibitem [{\citenamefont {Niu}\ \emph {et~al.}(2013)\citenamefont {Niu},
  \citenamefont {Niu}, \citenamefont {Paar}, \citenamefont {Vretenar},
  \citenamefont {Wang}, \citenamefont {Bai},\ and\ \citenamefont
  {Meng}}]{Niu2013_PRC88-034308}%
  \BibitemOpen
  \bibfield  {author} {\bibinfo {author} {\bibfnamefont {Y.~F.}\ \bibnamefont
  {Niu}}, \bibinfo {author} {\bibfnamefont {Z.~M.}\ \bibnamefont {Niu}},
  \bibinfo {author} {\bibfnamefont {N.}~\bibnamefont {Paar}}, \bibinfo {author}
  {\bibfnamefont {D.}~\bibnamefont {Vretenar}}, \bibinfo {author}
  {\bibfnamefont {G.~H.}\ \bibnamefont {Wang}}, \bibinfo {author}
  {\bibfnamefont {J.~S.}\ \bibnamefont {Bai}}, \ and\ \bibinfo {author}
  {\bibfnamefont {J.}~\bibnamefont {Meng}},\ }\href
  {http://link.aps.org/doi/10.1103/PhysRevC.88.034308} {\bibfield  {journal}
  {\bibinfo  {journal} {Phys. Rev. C}\ }\textbf {\bibinfo {volume} {88}},\
  \bibinfo {pages} {034308} (\bibinfo {year} {2013})}\BibitemShut {NoStop}%
\bibitem [{\citenamefont {Zhang}\ and\ \citenamefont
  {Niu}(2018)}]{Zhang2018_PRC97-054302}%
  \BibitemOpen
  \bibfield  {author} {\bibinfo {author} {\bibfnamefont {W.}~\bibnamefont
  {Zhang}}\ and\ \bibinfo {author} {\bibfnamefont {Y.~F.}\ \bibnamefont
  {Niu}},\ }\href {https://link.aps.org/doi/10.1103/PhysRevC.97.054302}
  {\bibfield  {journal} {\bibinfo  {journal} {Phys. Rev. C}\ }\textbf {\bibinfo
  {volume} {97}},\ \bibinfo {pages} {054302} (\bibinfo {year}
  {2018})}\BibitemShut {NoStop}%
\bibitem [{\citenamefont {Zhang}\ and\ \citenamefont
  {Niu}(2017)}]{Zhang2017_PRC96-054308}%
  \BibitemOpen
  \bibfield  {author} {\bibinfo {author} {\bibfnamefont {W.}~\bibnamefont
  {Zhang}}\ and\ \bibinfo {author} {\bibfnamefont {Y.~F.}\ \bibnamefont
  {Niu}},\ }\href {https://link.aps.org/doi/10.1103/PhysRevC.96.054308}
  {\bibfield  {journal} {\bibinfo  {journal} {Phys. Rev. C}\ }\textbf {\bibinfo
  {volume} {96}},\ \bibinfo {pages} {054308} (\bibinfo {year}
  {2017})}\BibitemShut {NoStop}%
\bibitem [{\citenamefont {Zhao}\ \emph
  {et~al.}(2019{\natexlab{a}})\citenamefont {Zhao}, \citenamefont {Xiang},
  \citenamefont {Li}, \citenamefont {Niksic}, \citenamefont {Vretenar},\ and\
  \citenamefont {Zhou}}]{Zhao2019_PRC99-054613}%
  \BibitemOpen
  \bibfield  {author} {\bibinfo {author} {\bibfnamefont {J.}~\bibnamefont
  {Zhao}}, \bibinfo {author} {\bibfnamefont {J.}~\bibnamefont {Xiang}},
  \bibinfo {author} {\bibfnamefont {Z.-P.}\ \bibnamefont {Li}}, \bibinfo
  {author} {\bibfnamefont {T.}~\bibnamefont {Niksic}}, \bibinfo {author}
  {\bibfnamefont {D.}~\bibnamefont {Vretenar}}, \ and\ \bibinfo {author}
  {\bibfnamefont {S.-G.}\ \bibnamefont {Zhou}},\ }\href
  {https://link.aps.org/doi/10.1103/PhysRevC.99.054613} {\bibfield  {journal}
  {\bibinfo  {journal} {Phys. Rev. C}\ }\textbf {\bibinfo {volume} {99}},\
  \bibinfo {pages} {054613} (\bibinfo {year} {2019}{\natexlab{a}})}\BibitemShut
  {NoStop}%
\bibitem [{\citenamefont {Zhao}\ \emph
  {et~al.}(2019{\natexlab{b}})\citenamefont {Zhao}, \citenamefont {Niksic},
  \citenamefont {Vretenar},\ and\ \citenamefont
  {Zhou}}]{Zhao2019_PRC99-014618}%
  \BibitemOpen
  \bibfield  {author} {\bibinfo {author} {\bibfnamefont {J.}~\bibnamefont
  {Zhao}}, \bibinfo {author} {\bibfnamefont {T.}~\bibnamefont {Niksic}},
  \bibinfo {author} {\bibfnamefont {D.}~\bibnamefont {Vretenar}}, \ and\
  \bibinfo {author} {\bibfnamefont {S.-G.}\ \bibnamefont {Zhou}},\ }\href
  {https://link.aps.org/doi/10.1103/PhysRevC.99.014618} {\bibfield  {journal}
  {\bibinfo  {journal} {Phys. Rev. C}\ }\textbf {\bibinfo {volume} {99}},\
  \bibinfo {pages} {014618} (\bibinfo {year} {2019}{\natexlab{b}})}\BibitemShut
  {NoStop}%
\bibitem [{\citenamefont {Goodman}(1981)}]{Goodman1981_NPA352-30}%
  \BibitemOpen
  \bibfield  {author} {\bibinfo {author} {\bibfnamefont {A.~L.}\ \bibnamefont
  {Goodman}},\ }\href
  {http://www.sciencedirect.com/science/article/pii/0375947481905571}
  {\bibfield  {journal} {\bibinfo  {journal} {Nucl. Phys. A}\ }\textbf
  {\bibinfo {volume} {352}},\ \bibinfo {pages} {30} (\bibinfo {year}
  {1981})}\BibitemShut {NoStop}%
\bibitem [{\citenamefont {Egido}\ \emph {et~al.}(1986)\citenamefont {Egido},
  \citenamefont {Ring},\ and\ \citenamefont {Mang}}]{Egido1986_NPA451-77}%
  \BibitemOpen
  \bibfield  {author} {\bibinfo {author} {\bibfnamefont {J.~L.}\ \bibnamefont
  {Egido}}, \bibinfo {author} {\bibfnamefont {P.}~\bibnamefont {Ring}}, \ and\
  \bibinfo {author} {\bibfnamefont {H.~J.}\ \bibnamefont {Mang}},\ }\href
  {http://www.sciencedirect.com/science/article/pii/0375947486902423}
  {\bibfield  {journal} {\bibinfo  {journal} {Nucl. Phys. A}\ }\textbf
  {\bibinfo {volume} {451}},\ \bibinfo {pages} {77} (\bibinfo {year}
  {1986})}\BibitemShut {NoStop}%
\bibitem [{\citenamefont {Lu}\ \emph {et~al.}(2012)\citenamefont {Lu},
  \citenamefont {Zhao},\ and\ \citenamefont {Zhou}}]{Lu2012_PRC85-011301R}%
  \BibitemOpen
  \bibfield  {author} {\bibinfo {author} {\bibfnamefont {B.-N.}\ \bibnamefont
  {Lu}}, \bibinfo {author} {\bibfnamefont {E.-G.}\ \bibnamefont {Zhao}}, \ and\
  \bibinfo {author} {\bibfnamefont {S.-G.}\ \bibnamefont {Zhou}},\ }\href
  {http://link.aps.org/doi/10.1103/PhysRevC.85.011301} {\bibfield  {journal}
  {\bibinfo  {journal} {Phys. Rev. C}\ }\textbf {\bibinfo {volume} {85}},\
  \bibinfo {pages} {011301(R)} (\bibinfo {year} {2012})}\BibitemShut {NoStop}%
\bibitem [{\citenamefont {Lu}\ \emph {et~al.}(2014)\citenamefont {Lu},
  \citenamefont {Zhao}, \citenamefont {Zhao},\ and\ \citenamefont
  {Zhou}}]{Lu2014_PRC89-014323}%
  \BibitemOpen
  \bibfield  {author} {\bibinfo {author} {\bibfnamefont {B.-N.}\ \bibnamefont
  {Lu}}, \bibinfo {author} {\bibfnamefont {J.}~\bibnamefont {Zhao}}, \bibinfo
  {author} {\bibfnamefont {E.-G.}\ \bibnamefont {Zhao}}, \ and\ \bibinfo
  {author} {\bibfnamefont {S.-G.}\ \bibnamefont {Zhou}},\ }\href
  {http://link.aps.org/doi/10.1103/PhysRevC.89.014323} {\bibfield  {journal}
  {\bibinfo  {journal} {Phys. Rev. C}\ }\textbf {\bibinfo {volume} {89}},\
  \bibinfo {pages} {014323} (\bibinfo {year} {2014})}\BibitemShut {NoStop}%
\bibitem [{\citenamefont {Zhao}\ \emph {et~al.}(2016)\citenamefont {Zhao},
  \citenamefont {Lu}, \citenamefont {Niksic}, \citenamefont {Vretenar},\ and\
  \citenamefont {Zhou}}]{Zhao2016_PRC93-044315}%
  \BibitemOpen
  \bibfield  {author} {\bibinfo {author} {\bibfnamefont {J.}~\bibnamefont
  {Zhao}}, \bibinfo {author} {\bibfnamefont {B.-N.}\ \bibnamefont {Lu}},
  \bibinfo {author} {\bibfnamefont {T.}~\bibnamefont {Niksic}}, \bibinfo
  {author} {\bibfnamefont {D.}~\bibnamefont {Vretenar}}, \ and\ \bibinfo
  {author} {\bibfnamefont {S.-G.}\ \bibnamefont {Zhou}},\ }\href
  {http://link.aps.org/doi/10.1103/PhysRevC.93.044315} {\bibfield  {journal}
  {\bibinfo  {journal} {Phys. Rev. C}\ }\textbf {\bibinfo {volume} {93}},\
  \bibinfo {pages} {044315} (\bibinfo {year} {2016})}\BibitemShut {NoStop}%
\bibitem [{\citenamefont {Nik\v{s}i\'c}\ \emph {et~al.}(2008)\citenamefont
  {Nik\v{s}i\'c}, \citenamefont {Vretenar},\ and\ \citenamefont
  {Ring}}]{Niksic2008_PRC78-034318}%
  \BibitemOpen
  \bibfield  {author} {\bibinfo {author} {\bibfnamefont {T.}~\bibnamefont
  {Nik\v{s}i\'c}}, \bibinfo {author} {\bibfnamefont {D.}~\bibnamefont
  {Vretenar}}, \ and\ \bibinfo {author} {\bibfnamefont {P.}~\bibnamefont
  {Ring}},\ }\href {http://link.aps.org/doi/10.1103/PhysRevC.78.034318}
  {\bibfield  {journal} {\bibinfo  {journal} {Phys. Rev. C}\ }\textbf {\bibinfo
  {volume} {78}},\ \bibinfo {pages} {034318} (\bibinfo {year}
  {2008})}\BibitemShut {NoStop}%
\bibitem [{\citenamefont {Tian}\ \emph {et~al.}(2009)\citenamefont {Tian},
  \citenamefont {Ma},\ and\ \citenamefont {Ring}}]{Tian2009_PLB676-44}%
  \BibitemOpen
  \bibfield  {author} {\bibinfo {author} {\bibfnamefont {Y.}~\bibnamefont
  {Tian}}, \bibinfo {author} {\bibfnamefont {Z.~Y.}\ \bibnamefont {Ma}}, \ and\
  \bibinfo {author} {\bibfnamefont {P.}~\bibnamefont {Ring}},\ }\href
  {http://www.sciencedirect.com/science/article/pii/S0370269309004912}
  {\bibfield  {journal} {\bibinfo  {journal} {Phys. Lett. B}\ }\textbf
  {\bibinfo {volume} {676}},\ \bibinfo {pages} {44} (\bibinfo {year}
  {2009})}\BibitemShut {NoStop}%
\bibitem [{\citenamefont {Berger}\ \emph {et~al.}(1991)\citenamefont {Berger},
  \citenamefont {Girod},\ and\ \citenamefont {Gogny}}]{Berger1991_CPC63-365}%
  \BibitemOpen
  \bibfield  {author} {\bibinfo {author} {\bibfnamefont {J.}~\bibnamefont
  {Berger}}, \bibinfo {author} {\bibfnamefont {M.}~\bibnamefont {Girod}}, \
  and\ \bibinfo {author} {\bibfnamefont {D.}~\bibnamefont {Gogny}},\ }\href
  {http://www.sciencedirect.com/science/article/pii/001046559190263K}
  {\bibfield  {journal} {\bibinfo  {journal} {Comput. Phys. Commun.}\ }\textbf
  {\bibinfo {volume} {63}},\ \bibinfo {pages} {365} (\bibinfo {year}
  {1991})}\BibitemShut {NoStop}%
\bibitem [{\citenamefont {Nik\v{s}i\'c}\ \emph {et~al.}(2009)\citenamefont
  {Nik\v{s}i\'c}, \citenamefont {Li}, \citenamefont {Vretenar}, \citenamefont
  {Pr\'ochniak}, \citenamefont {Meng},\ and\ \citenamefont
  {Ring}}]{Niksic2009_PRC79-034303}%
  \BibitemOpen
  \bibfield  {author} {\bibinfo {author} {\bibfnamefont {T.}~\bibnamefont
  {Nik\v{s}i\'c}}, \bibinfo {author} {\bibfnamefont {Z.~P.}\ \bibnamefont
  {Li}}, \bibinfo {author} {\bibfnamefont {D.}~\bibnamefont {Vretenar}},
  \bibinfo {author} {\bibfnamefont {L.}~\bibnamefont {Pr\'ochniak}}, \bibinfo
  {author} {\bibfnamefont {J.}~\bibnamefont {Meng}}, \ and\ \bibinfo {author}
  {\bibfnamefont {P.}~\bibnamefont {Ring}},\ }\href
  {http://link.aps.org/doi/10.1103/PhysRevC.79.034303} {\bibfield  {journal}
  {\bibinfo  {journal} {Phys. Rev. C}\ }\textbf {\bibinfo {volume} {79}},\
  \bibinfo {pages} {034303} (\bibinfo {year} {2009})}\BibitemShut {NoStop}%
\bibitem [{\citenamefont {Li}\ \emph {et~al.}(2013)\citenamefont {Li},
  \citenamefont {Song}, \citenamefont {Yao}, \citenamefont {Vretenar},\ and\
  \citenamefont {Meng}}]{Li2013_PLB726-866}%
  \BibitemOpen
  \bibfield  {author} {\bibinfo {author} {\bibfnamefont {Z.}~\bibnamefont
  {Li}}, \bibinfo {author} {\bibfnamefont {B.}~\bibnamefont {Song}}, \bibinfo
  {author} {\bibfnamefont {J.}~\bibnamefont {Yao}}, \bibinfo {author}
  {\bibfnamefont {D.}~\bibnamefont {Vretenar}}, \ and\ \bibinfo {author}
  {\bibfnamefont {J.}~\bibnamefont {Meng}},\ }\href
  {http://www.sciencedirect.com/science/article/pii/S0370269313007648}
  {\bibfield  {journal} {\bibinfo  {journal} {Phys. Lett. B}\ }\textbf
  {\bibinfo {volume} {726}},\ \bibinfo {pages} {866} (\bibinfo {year}
  {2013})}\BibitemShut {NoStop}%
\bibitem [{\citenamefont {Zhao}\ \emph {et~al.}(2015)\citenamefont {Zhao},
  \citenamefont {Lu}, \citenamefont {Niksic},\ and\ \citenamefont
  {Vretenar}}]{Zhao2015_PRC92-064315}%
  \BibitemOpen
  \bibfield  {author} {\bibinfo {author} {\bibfnamefont {J.}~\bibnamefont
  {Zhao}}, \bibinfo {author} {\bibfnamefont {B.-N.}\ \bibnamefont {Lu}},
  \bibinfo {author} {\bibfnamefont {T.}~\bibnamefont {Niksic}}, \ and\ \bibinfo
  {author} {\bibfnamefont {D.}~\bibnamefont {Vretenar}},\ }\href
  {http://link.aps.org/doi/10.1103/PhysRevC.92.064315} {\bibfield  {journal}
  {\bibinfo  {journal} {Phys. Rev. C}\ }\textbf {\bibinfo {volume} {92}},\
  \bibinfo {pages} {064315} (\bibinfo {year} {2015})}\BibitemShut {NoStop}%
\bibitem [{\citenamefont {Sadhukhan}\ \emph {et~al.}(2013)\citenamefont
  {Sadhukhan}, \citenamefont {Mazurek}, \citenamefont {Baran}, \citenamefont
  {Dobaczewski}, \citenamefont {Nazarewicz},\ and\ \citenamefont
  {Sheikh}}]{Sadhukhan2013_PRC88-064314}%
  \BibitemOpen
  \bibfield  {author} {\bibinfo {author} {\bibfnamefont {J.}~\bibnamefont
  {Sadhukhan}}, \bibinfo {author} {\bibfnamefont {K.}~\bibnamefont {Mazurek}},
  \bibinfo {author} {\bibfnamefont {A.}~\bibnamefont {Baran}}, \bibinfo
  {author} {\bibfnamefont {J.}~\bibnamefont {Dobaczewski}}, \bibinfo {author}
  {\bibfnamefont {W.}~\bibnamefont {Nazarewicz}}, \ and\ \bibinfo {author}
  {\bibfnamefont {J.~A.}\ \bibnamefont {Sheikh}},\ }\href
  {http://link.aps.org/doi/10.1103/PhysRevC.88.064314} {\bibfield  {journal}
  {\bibinfo  {journal} {Phys. Rev. C}\ }\textbf {\bibinfo {volume} {88}},\
  \bibinfo {pages} {064314} (\bibinfo {year} {2013})}\BibitemShut {NoStop}%
\bibitem [{\citenamefont {Sadhukhan}\ \emph {et~al.}(2014)\citenamefont
  {Sadhukhan}, \citenamefont {Dobaczewski}, \citenamefont {Nazarewicz},
  \citenamefont {Sheikh},\ and\ \citenamefont
  {Baran}}]{Sadhukhan2014_PRC90-061304}%
  \BibitemOpen
  \bibfield  {author} {\bibinfo {author} {\bibfnamefont {J.}~\bibnamefont
  {Sadhukhan}}, \bibinfo {author} {\bibfnamefont {J.}~\bibnamefont
  {Dobaczewski}}, \bibinfo {author} {\bibfnamefont {W.}~\bibnamefont
  {Nazarewicz}}, \bibinfo {author} {\bibfnamefont {J.~A.}\ \bibnamefont
  {Sheikh}}, \ and\ \bibinfo {author} {\bibfnamefont {A.}~\bibnamefont
  {Baran}},\ }\href {http://link.aps.org/doi/10.1103/PhysRevC.90.061304}
  {\bibfield  {journal} {\bibinfo  {journal} {Phys. Rev. C}\ }\textbf {\bibinfo
  {volume} {90}},\ \bibinfo {pages} {061304} (\bibinfo {year}
  {2014})}\BibitemShut {NoStop}%
\bibitem [{\citenamefont {Baran}\ \emph {et~al.}(2007)\citenamefont {Baran},
  \citenamefont {Staszczak}, \citenamefont {Dobaczewski},\ and\ \citenamefont
  {Nazarewicz}}]{Baran2007_IJMPE16-443}%
  \BibitemOpen
  \bibfield  {author} {\bibinfo {author} {\bibfnamefont {A.}~\bibnamefont
  {Baran}}, \bibinfo {author} {\bibfnamefont {A.}~\bibnamefont {Staszczak}},
  \bibinfo {author} {\bibfnamefont {J.}~\bibnamefont {Dobaczewski}}, \ and\
  \bibinfo {author} {\bibfnamefont {W.}~\bibnamefont {Nazarewicz}},\ }\href
  {http://dx.doi.org/10.1142/S0218301307005879} {\bibfield  {journal} {\bibinfo
   {journal} {Int. J. Mod. Phys. E}\ }\textbf {\bibinfo {volume} {16}},\
  \bibinfo {pages} {443} (\bibinfo {year} {2007})}\BibitemShut {NoStop}%
\bibitem [{\citenamefont {Staszczak}\ \emph {et~al.}(2013)\citenamefont
  {Staszczak}, \citenamefont {Baran},\ and\ \citenamefont
  {Nazarewicz}}]{Staszczak2013_PRC87-024320}%
  \BibitemOpen
  \bibfield  {author} {\bibinfo {author} {\bibfnamefont {A.}~\bibnamefont
  {Staszczak}}, \bibinfo {author} {\bibfnamefont {A.}~\bibnamefont {Baran}}, \
  and\ \bibinfo {author} {\bibfnamefont {W.}~\bibnamefont {Nazarewicz}},\
  }\href {http://link.aps.org/doi/10.1103/PhysRevC.87.024320} {\bibfield
  {journal} {\bibinfo  {journal} {Phys. Rev. C}\ }\textbf {\bibinfo {volume}
  {87}},\ \bibinfo {pages} {024320} (\bibinfo {year} {2013})}\BibitemShut
  {NoStop}%
\bibitem [{\citenamefont {Zhao}\ \emph {et~al.}(2017)\citenamefont {Zhao},
  \citenamefont {Lu}, \citenamefont {Zhao},\ and\ \citenamefont
  {Zhou}}]{Zhao2017_PRC95-014320}%
  \BibitemOpen
  \bibfield  {author} {\bibinfo {author} {\bibfnamefont {J.}~\bibnamefont
  {Zhao}}, \bibinfo {author} {\bibfnamefont {B.-N.}\ \bibnamefont {Lu}},
  \bibinfo {author} {\bibfnamefont {E.-G.}\ \bibnamefont {Zhao}}, \ and\
  \bibinfo {author} {\bibfnamefont {S.-G.}\ \bibnamefont {Zhou}},\ }\href
  {\doibase 10.1103/PhysRevC.95.014320} {\bibfield  {journal} {\bibinfo
  {journal} {Phys. Rev. C}\ }\textbf {\bibinfo {volume} {95}},\ \bibinfo
  {pages} {014320} (\bibinfo {year} {2017})}\BibitemShut {NoStop}%
\bibitem [{\citenamefont {Fransen}\ \emph {et~al.}(2003)\citenamefont
  {Fransen}, \citenamefont {Pietralla}, \citenamefont {Ammar}, \citenamefont
  {Bandyopadhyay}, \citenamefont {Boukharouba}, \citenamefont {von Brentano},
  \citenamefont {Dewald}, \citenamefont {Gableske}, \citenamefont {Gade},
  \citenamefont {Jolie}, \citenamefont {Kneissl}, \citenamefont {Lesher},
  \citenamefont {Lisetskiy}, \citenamefont {McEllistrem}, \citenamefont
  {Merrick}, \citenamefont {Pitz}, \citenamefont {Warr}, \citenamefont
  {Werner},\ and\ \citenamefont {Yates}}]{Fransen2003_PRC67-024307}%
  \BibitemOpen
  \bibfield  {author} {\bibinfo {author} {\bibfnamefont {C.}~\bibnamefont
  {Fransen}}, \bibinfo {author} {\bibfnamefont {N.}~\bibnamefont {Pietralla}},
  \bibinfo {author} {\bibfnamefont {Z.}~\bibnamefont {Ammar}}, \bibinfo
  {author} {\bibfnamefont {D.}~\bibnamefont {Bandyopadhyay}}, \bibinfo {author}
  {\bibfnamefont {N.}~\bibnamefont {Boukharouba}}, \bibinfo {author}
  {\bibfnamefont {P.}~\bibnamefont {von Brentano}}, \bibinfo {author}
  {\bibfnamefont {A.}~\bibnamefont {Dewald}}, \bibinfo {author} {\bibfnamefont
  {J.}~\bibnamefont {Gableske}}, \bibinfo {author} {\bibfnamefont
  {A.}~\bibnamefont {Gade}}, \bibinfo {author} {\bibfnamefont {J.}~\bibnamefont
  {Jolie}}, \bibinfo {author} {\bibfnamefont {U.}~\bibnamefont {Kneissl}},
  \bibinfo {author} {\bibfnamefont {S.~R.}\ \bibnamefont {Lesher}}, \bibinfo
  {author} {\bibfnamefont {A.~F.}\ \bibnamefont {Lisetskiy}}, \bibinfo {author}
  {\bibfnamefont {M.~T.}\ \bibnamefont {McEllistrem}}, \bibinfo {author}
  {\bibfnamefont {M.}~\bibnamefont {Merrick}}, \bibinfo {author} {\bibfnamefont
  {H.~H.}\ \bibnamefont {Pitz}}, \bibinfo {author} {\bibfnamefont
  {N.}~\bibnamefont {Warr}}, \bibinfo {author} {\bibfnamefont {V.}~\bibnamefont
  {Werner}}, \ and\ \bibinfo {author} {\bibfnamefont {S.~W.}\ \bibnamefont
  {Yates}},\ }\href {\doibase 10.1103/physrevc.67.024307} {\bibfield  {journal}
  {\bibinfo  {journal} {Phys. Rev. C}\ }\textbf {\bibinfo {volume} {67}},\
  \bibinfo {pages} {024307} (\bibinfo {year} {2003})}\BibitemShut {NoStop}%
\bibitem [{\citenamefont {Mu}\ and\ \citenamefont
  {Zhang}(2018)}]{Mu2018_SCPMA61-012011}%
  \BibitemOpen
  \bibfield  {author} {\bibinfo {author} {\bibfnamefont {C.}~\bibnamefont
  {Mu}}\ and\ \bibinfo {author} {\bibfnamefont {D.}~\bibnamefont {Zhang}},\
  }\href {\doibase 10.1007/s11433-017-9106-5} {\bibfield  {journal} {\bibinfo
  {journal} {Sci. China-Phys. Mech. Astron.}\ }\textbf {\bibinfo {volume}
  {61}},\ \bibinfo {pages} {012011} (\bibinfo {year} {2018})}\BibitemShut
  {NoStop}%
\bibitem [{\citenamefont {Utsunomiya}\ \emph {et~al.}(2013)\citenamefont
  {Utsunomiya}, \citenamefont {Goriely}, \citenamefont {Kondo}, \citenamefont
  {Iwamoto}, \citenamefont {Akimune}, \citenamefont {Yamagata}, \citenamefont
  {Toyokawa}, \citenamefont {Harada}, \citenamefont {Kitatani}, \citenamefont
  {Lui}, \citenamefont {Larsen}, \citenamefont {Guttormsen}, \citenamefont
  {Koehler}, \citenamefont {Hilaire}, \citenamefont {P{\'{e}}ru}, \citenamefont
  {Martini},\ and\ \citenamefont {Koning}}]{Utsunomiya2013_PRC88-015805}%
  \BibitemOpen
  \bibfield  {author} {\bibinfo {author} {\bibfnamefont {H.}~\bibnamefont
  {Utsunomiya}}, \bibinfo {author} {\bibfnamefont {S.}~\bibnamefont {Goriely}},
  \bibinfo {author} {\bibfnamefont {T.}~\bibnamefont {Kondo}}, \bibinfo
  {author} {\bibfnamefont {C.}~\bibnamefont {Iwamoto}}, \bibinfo {author}
  {\bibfnamefont {H.}~\bibnamefont {Akimune}}, \bibinfo {author} {\bibfnamefont
  {T.}~\bibnamefont {Yamagata}}, \bibinfo {author} {\bibfnamefont
  {H.}~\bibnamefont {Toyokawa}}, \bibinfo {author} {\bibfnamefont
  {H.}~\bibnamefont {Harada}}, \bibinfo {author} {\bibfnamefont
  {F.}~\bibnamefont {Kitatani}}, \bibinfo {author} {\bibfnamefont {Y.-W.}\
  \bibnamefont {Lui}}, \bibinfo {author} {\bibfnamefont {A.~C.}\ \bibnamefont
  {Larsen}}, \bibinfo {author} {\bibfnamefont {M.}~\bibnamefont {Guttormsen}},
  \bibinfo {author} {\bibfnamefont {P.~E.}\ \bibnamefont {Koehler}}, \bibinfo
  {author} {\bibfnamefont {S.}~\bibnamefont {Hilaire}}, \bibinfo {author}
  {\bibfnamefont {S.}~\bibnamefont {P{\'{e}}ru}}, \bibinfo {author}
  {\bibfnamefont {M.}~\bibnamefont {Martini}}, \ and\ \bibinfo {author}
  {\bibfnamefont {A.~J.}\ \bibnamefont {Koning}},\ }\href {\doibase
  10.1103/physrevc.88.015805} {\bibfield  {journal} {\bibinfo  {journal} {Phys.
  Rev. C}\ }\textbf {\bibinfo {volume} {88}},\ \bibinfo {pages} {015805}
  (\bibinfo {year} {2013})}\BibitemShut {NoStop}%
\bibitem [{\citenamefont {Eriksen}\ \emph {et~al.}(2014)\citenamefont
  {Eriksen}, \citenamefont {Nyhus}, \citenamefont {Guttormsen}, \citenamefont
  {G{\"o}rgen}, \citenamefont {Larsen}, \citenamefont {Renstr{\o}m},
  \citenamefont {Ruud}, \citenamefont {Siem}, \citenamefont {Toft},
  \citenamefont {Tveten},\ and\ \citenamefont
  {Wilson}}]{Eriksen2014_PRC90-044311}%
  \BibitemOpen
  \bibfield  {author} {\bibinfo {author} {\bibfnamefont {T.~K.}\ \bibnamefont
  {Eriksen}}, \bibinfo {author} {\bibfnamefont {H.~T.}\ \bibnamefont {Nyhus}},
  \bibinfo {author} {\bibfnamefont {M.}~\bibnamefont {Guttormsen}}, \bibinfo
  {author} {\bibfnamefont {A.}~\bibnamefont {G{\"o}rgen}}, \bibinfo {author}
  {\bibfnamefont {A.~C.}\ \bibnamefont {Larsen}}, \bibinfo {author}
  {\bibfnamefont {T.}~\bibnamefont {Renstr{\o}m}}, \bibinfo {author}
  {\bibfnamefont {I.~E.}\ \bibnamefont {Ruud}}, \bibinfo {author}
  {\bibfnamefont {S.}~\bibnamefont {Siem}}, \bibinfo {author} {\bibfnamefont
  {H.~K.}\ \bibnamefont {Toft}}, \bibinfo {author} {\bibfnamefont {G.~M.}\
  \bibnamefont {Tveten}}, \ and\ \bibinfo {author} {\bibfnamefont {J.~N.}\
  \bibnamefont {Wilson}},\ }\href {\doibase 10.1103/physrevc.90.044311}
  {\bibfield  {journal} {\bibinfo  {journal} {Phys. Rev. C}\ }\textbf {\bibinfo
  {volume} {90}},\ \bibinfo {pages} {044311} (\bibinfo {year}
  {2014})}\BibitemShut {NoStop}%
\bibitem [{\citenamefont {Larsen}\ \emph {et~al.}(2013)\citenamefont {Larsen},
  \citenamefont {Ruud}, \citenamefont {B{\"u}rger}, \citenamefont {Goriely},
  \citenamefont {Guttormsen}, \citenamefont {G{\"o}rgen}, \citenamefont
  {Hagen}, \citenamefont {Harissopulos}, \citenamefont {Nyhus}, \citenamefont
  {Renstr{\o}m}, \citenamefont {Schiller}, \citenamefont {Siem}, \citenamefont
  {Tveten}, \citenamefont {Voinov},\ and\ \citenamefont
  {Wiedeking}}]{Larsen2013_PRC87-014319}%
  \BibitemOpen
  \bibfield  {author} {\bibinfo {author} {\bibfnamefont {A.~C.}\ \bibnamefont
  {Larsen}}, \bibinfo {author} {\bibfnamefont {I.~E.}\ \bibnamefont {Ruud}},
  \bibinfo {author} {\bibfnamefont {A.}~\bibnamefont {B{\"u}rger}}, \bibinfo
  {author} {\bibfnamefont {S.}~\bibnamefont {Goriely}}, \bibinfo {author}
  {\bibfnamefont {M.}~\bibnamefont {Guttormsen}}, \bibinfo {author}
  {\bibfnamefont {A.}~\bibnamefont {G{\"o}rgen}}, \bibinfo {author}
  {\bibfnamefont {T.~W.}\ \bibnamefont {Hagen}}, \bibinfo {author}
  {\bibfnamefont {S.}~\bibnamefont {Harissopulos}}, \bibinfo {author}
  {\bibfnamefont {H.~T.}\ \bibnamefont {Nyhus}}, \bibinfo {author}
  {\bibfnamefont {T.}~\bibnamefont {Renstr{\o}m}}, \bibinfo {author}
  {\bibfnamefont {A.}~\bibnamefont {Schiller}}, \bibinfo {author}
  {\bibfnamefont {S.}~\bibnamefont {Siem}}, \bibinfo {author} {\bibfnamefont
  {G.~M.}\ \bibnamefont {Tveten}}, \bibinfo {author} {\bibfnamefont
  {A.}~\bibnamefont {Voinov}}, \ and\ \bibinfo {author} {\bibfnamefont
  {M.}~\bibnamefont {Wiedeking}},\ }\href {\doibase 10.1103/physrevc.87.014319}
  {\bibfield  {journal} {\bibinfo  {journal} {Phys. Rev. C}\ }\textbf {\bibinfo
  {volume} {87}},\ \bibinfo {pages} {014319} (\bibinfo {year}
  {2013})}\BibitemShut {NoStop}%
\bibitem [{\citenamefont {Guttormsen}\ \emph {et~al.}(2003)\citenamefont
  {Guttormsen}, \citenamefont {Bagheri}, \citenamefont {Chankova},
  \citenamefont {Rekstad}, \citenamefont {Siem}, \citenamefont {Schiller},\
  and\ \citenamefont {Voinov}}]{Guttormsen2003_PRC68-064306}%
  \BibitemOpen
  \bibfield  {author} {\bibinfo {author} {\bibfnamefont {M.}~\bibnamefont
  {Guttormsen}}, \bibinfo {author} {\bibfnamefont {A.}~\bibnamefont {Bagheri}},
  \bibinfo {author} {\bibfnamefont {R.}~\bibnamefont {Chankova}}, \bibinfo
  {author} {\bibfnamefont {J.}~\bibnamefont {Rekstad}}, \bibinfo {author}
  {\bibfnamefont {S.}~\bibnamefont {Siem}}, \bibinfo {author} {\bibfnamefont
  {A.}~\bibnamefont {Schiller}}, \ and\ \bibinfo {author} {\bibfnamefont
  {A.}~\bibnamefont {Voinov}},\ }\href {\doibase 10.1103/physrevc.68.064306}
  {\bibfield  {journal} {\bibinfo  {journal} {Phys. Rev. C}\ }\textbf {\bibinfo
  {volume} {68}},\ \bibinfo {pages} {064306} (\bibinfo {year}
  {2003})}\BibitemShut {NoStop}%
\bibitem [{\citenamefont {Schiller}\ \emph {et~al.}(2001)\citenamefont
  {Schiller}, \citenamefont {Bjerve}, \citenamefont {Guttormsen}, \citenamefont
  {Hjorth-Jensen}, \citenamefont {Ingebretsen}, \citenamefont {Melby},
  \citenamefont {Messelt}, \citenamefont {Rekstad}, \citenamefont {Siem},\ and\
  \citenamefont {{\O}deg{\aa}rd}}]{Schiller2001_PRC63-021306}%
  \BibitemOpen
  \bibfield  {author} {\bibinfo {author} {\bibfnamefont {A.}~\bibnamefont
  {Schiller}}, \bibinfo {author} {\bibfnamefont {A.}~\bibnamefont {Bjerve}},
  \bibinfo {author} {\bibfnamefont {M.}~\bibnamefont {Guttormsen}}, \bibinfo
  {author} {\bibfnamefont {M.}~\bibnamefont {Hjorth-Jensen}}, \bibinfo {author}
  {\bibfnamefont {F.}~\bibnamefont {Ingebretsen}}, \bibinfo {author}
  {\bibfnamefont {E.}~\bibnamefont {Melby}}, \bibinfo {author} {\bibfnamefont
  {S.}~\bibnamefont {Messelt}}, \bibinfo {author} {\bibfnamefont
  {J.}~\bibnamefont {Rekstad}}, \bibinfo {author} {\bibfnamefont
  {S.}~\bibnamefont {Siem}}, \ and\ \bibinfo {author} {\bibfnamefont {S.~W.}\
  \bibnamefont {{\O}deg{\aa}rd}},\ }\href {\doibase 10.1103/physrevc.63.021306}
  {\bibfield  {journal} {\bibinfo  {journal} {Phys. Rev. C}\ }\textbf {\bibinfo
  {volume} {63}},\ \bibinfo {pages} {021306} (\bibinfo {year}
  {2001})}\BibitemShut {NoStop}%
\bibitem [{\citenamefont {Nyhus}\ \emph {et~al.}(2012)\citenamefont {Nyhus},
  \citenamefont {Siem}, \citenamefont {Guttormsen}, \citenamefont {Larsen},
  \citenamefont {B{\"u}rger}, \citenamefont {Syed}, \citenamefont {Toft},
  \citenamefont {Tveten},\ and\ \citenamefont
  {Voinov}}]{Nyhus2012_PRC85-014323}%
  \BibitemOpen
  \bibfield  {author} {\bibinfo {author} {\bibfnamefont {H.~T.}\ \bibnamefont
  {Nyhus}}, \bibinfo {author} {\bibfnamefont {S.}~\bibnamefont {Siem}},
  \bibinfo {author} {\bibfnamefont {M.}~\bibnamefont {Guttormsen}}, \bibinfo
  {author} {\bibfnamefont {A.~C.}\ \bibnamefont {Larsen}}, \bibinfo {author}
  {\bibfnamefont {A.}~\bibnamefont {B{\"u}rger}}, \bibinfo {author}
  {\bibfnamefont {N.~U.~H.}\ \bibnamefont {Syed}}, \bibinfo {author}
  {\bibfnamefont {H.~K.}\ \bibnamefont {Toft}}, \bibinfo {author}
  {\bibfnamefont {G.~M.}\ \bibnamefont {Tveten}}, \ and\ \bibinfo {author}
  {\bibfnamefont {A.}~\bibnamefont {Voinov}},\ }\href {\doibase
  10.1103/physrevc.85.014323} {\bibfield  {journal} {\bibinfo  {journal} {Phys.
  Rev. C}\ }\textbf {\bibinfo {volume} {85}},\ \bibinfo {pages} {014323}
  (\bibinfo {year} {2012})}\BibitemShut {NoStop}%
\bibitem [{\citenamefont {Melby}\ \emph {et~al.}(2001)\citenamefont {Melby},
  \citenamefont {Guttormsen}, \citenamefont {Rekstad}, \citenamefont
  {Schiller}, \citenamefont {Siem},\ and\ \citenamefont
  {Voinov}}]{Melby2001_PRC63-044309}%
  \BibitemOpen
  \bibfield  {author} {\bibinfo {author} {\bibfnamefont {E.}~\bibnamefont
  {Melby}}, \bibinfo {author} {\bibfnamefont {M.}~\bibnamefont {Guttormsen}},
  \bibinfo {author} {\bibfnamefont {J.}~\bibnamefont {Rekstad}}, \bibinfo
  {author} {\bibfnamefont {A.}~\bibnamefont {Schiller}}, \bibinfo {author}
  {\bibfnamefont {S.}~\bibnamefont {Siem}}, \ and\ \bibinfo {author}
  {\bibfnamefont {A.}~\bibnamefont {Voinov}},\ }\href {\doibase
  10.1103/physrevc.63.044309} {\bibfield  {journal} {\bibinfo  {journal} {Phys.
  Rev. C}\ }\textbf {\bibinfo {volume} {63}},\ \bibinfo {pages} {044309}
  (\bibinfo {year} {2001})}\BibitemShut {NoStop}%
\bibitem [{\citenamefont {Agvaanluvsan}\ \emph {et~al.}(2004)\citenamefont
  {Agvaanluvsan}, \citenamefont {Schiller}, \citenamefont {Becker},
  \citenamefont {Bernstein}, \citenamefont {Garrett}, \citenamefont
  {Guttormsen}, \citenamefont {Mitchell}, \citenamefont {Rekstad},
  \citenamefont {Siem}, \citenamefont {Voinov},\ and\ \citenamefont
  {Younes}}]{Agvaanluvsan2004_PRC70-054611}%
  \BibitemOpen
  \bibfield  {author} {\bibinfo {author} {\bibfnamefont {U.}~\bibnamefont
  {Agvaanluvsan}}, \bibinfo {author} {\bibfnamefont {A.}~\bibnamefont
  {Schiller}}, \bibinfo {author} {\bibfnamefont {J.~A.}\ \bibnamefont
  {Becker}}, \bibinfo {author} {\bibfnamefont {L.~A.}\ \bibnamefont
  {Bernstein}}, \bibinfo {author} {\bibfnamefont {P.~E.}\ \bibnamefont
  {Garrett}}, \bibinfo {author} {\bibfnamefont {M.}~\bibnamefont {Guttormsen}},
  \bibinfo {author} {\bibfnamefont {G.~E.}\ \bibnamefont {Mitchell}}, \bibinfo
  {author} {\bibfnamefont {J.}~\bibnamefont {Rekstad}}, \bibinfo {author}
  {\bibfnamefont {S.}~\bibnamefont {Siem}}, \bibinfo {author} {\bibfnamefont
  {A.}~\bibnamefont {Voinov}}, \ and\ \bibinfo {author} {\bibfnamefont
  {W.}~\bibnamefont {Younes}},\ }\href {\doibase 10.1103/physrevc.70.054611}
  {\bibfield  {journal} {\bibinfo  {journal} {Phys. Rev. C}\ }\textbf {\bibinfo
  {volume} {70}},\ \bibinfo {pages} {054611} (\bibinfo {year}
  {2004})}\BibitemShut {NoStop}%
\end{thebibliography}

%

\end{document}